\documentclass[sigconf]{acmart}

\copyrightyear{2024}
\acmYear{2024}
\setcopyright{rightsretained}
\acmConference[MOBIHOC '24]{The Twenty-fifth International Symposium on Theory, Algorithmic Foundations, and Protocol Design for Mobile Networks and Mobile Computing}{October 14--17, 2024}{Athens, Greece}
\acmBooktitle{The Twenty-fifth International Symposium on Theory, Algorithmic Foundations, and Protocol Design for Mobile Networks and Mobile Computing (MOBIHOC '24), October 14--17, 2024, Athens, Greece}\acmDOI{10.1145/3641512.3686358}
\acmISBN{979-8-4007-0521-2/24/10}

\usepackage{makecell}
\usepackage{amsmath,amsfonts}

\usepackage{amssymb}
\usepackage{utfsym}
\allowdisplaybreaks
\usepackage{nccmath}
\usepackage{algorithm}
\usepackage{algorithmic}
\usepackage{graphicx}
\usepackage{textcomp}
\usepackage{xcolor}
\usepackage{float}
\usepackage{subfigure}
\usepackage{subeqnarray}
\usepackage{adjustbox}
\PassOptionsToPackage{hyphens}{url}\usepackage{hyperref}
\usepackage[justification=centering]{caption}
 
\usepackage{graphicx}  

\usepackage{caption}
 
\DeclareCaptionLabelFormat{lc}{{#1}~#2}
\usepackage{pifont} 

\makeatletter
\def\endthebibliography{%
  \def\@noitemerr{\@latex@warning{Empty `thebibliography' environment}}%
  \endlist
}
\makeatother

 \usepackage{epsfig,endnotes}

\usepackage{floatpag,enumitem}

\usepackage{relsize}
 
 \usepackage{tikz}

\usepackage{multirow}
\usepackage{setspace} 
\usepackage{mathrsfs}
\usepackage{bbm}
\usepackage{pifont}
\usepackage{fancyhdr}
\usepackage{amsfonts}
\usepackage{amsmath, amsthm}

\usepackage{tabularx}
\usepackage{listings}
\usepackage{graphicx}
\usepackage{amssymb,booktabs}
\usepackage{array}
\usepackage{cases}

 \usepackage{supertabular}
\usepackage{mathrsfs}

\usepackage{psfrag}
\usepackage{bbding}

\usepackage{float}

\usepackage{amsbsy}
\usepackage{subcaption}
\makeatletter
\providecommand{\leftsquigarrow}{%
  \mathrel{\mathpalette\reflect@squig\relax}%
}
\newcommand{\reflect@squig}[2]{%
  \reflectbox{$\m@th#1\rightsquigarrow$}%
}
\makeatother

\usepackage{algorithm}
 \usepackage{algorithmic}

\makeatletter
\newcommand{\newalgname}[1]{%
  \renewcommand{\ALG@name}{#1}%
}

\newcommand {\C} {{\rm I\kern-5.5pt C}}

\def\centerhack#1{\hbox to 0pt{\hss\footnotesize #1\hss}}
\def\centerhackn#1{\hbox to 0pt{\hss #1\hss}}
\def\dchack#1{\vbox to 0pt{\vss{\hbox to 0pt{\hss#1\hss}}\vss}}

\newcommand{\argmin}{\operatornamewithlimits{argmin}}
\usepackage{etoolbox}
\AtBeginEnvironment{align}{\setcounter{subeqn}{0}}
\newcounter{subeqn} %

\usetikzlibrary{positioning}

\newcounter{mysub}

\newtheorem{defn}{Definition}

\newtheorem{lem}{Lemma}

\newtheorem{thm}{Theorem}

\newtheorem{prop}{Proposition}
\newtheorem*{proposition1.1}{Proposition 1.1}
\newtheorem*{proposition1.2}{Proposition 1.2}
\newtheorem*{proposition1.3}{Proposition 1.3}
\newtheorem*{proposition2.1}{Proposition 2.1}
\newtheorem*{proposition2.2}{Proposition 2.2}
\usepackage{subfigure}

\makeatletter
\newcommand{\linebreakand}{%
  \end{@IEEEauthorhalign}
  \hfill\mbox{}\par
  \mbox{}\hfill\begin{@IEEEauthorhalign}
}
\makeatother

\begin{document}

\title{Resource Allocation for Stable LLM Training in Mobile Edge Computing}


\author{Chang Liu}
\affiliation{%
  \institution{Graduate College\\Nanyang Technological University}
  \country{Singapore}
  }
\email{liuc0063@e.ntu.edu.sg}

\author{Jun Zhao}
\affiliation{%
 \institution{College of Computing and Data Science\\Nanyang Technological University}
 \country{Singapore}
 }
 \email{junzhao@ntu.edu.sg}
 
\thanks{Corresponding author: Jun Zhao\\Chang Liu is a PhD student supervised by Jun Zhao.}

\renewcommand{\shortauthors}{Liu and Zhao}

\begin{abstract}
As mobile devices increasingly become focal points for advanced applications, edge computing presents a viable solution to their inherent computational limitations, particularly in deploying large language models (LLMs).
However, despite the advancements in edge computing, significant challenges remain in efficient training and deploying LLMs due to the computational demands and data privacy concerns associated with these models.
This paper explores a collaborative training framework that integrates mobile users with edge servers to optimize resource allocation, thereby enhancing both performance and efficiency.
Our approach leverages parameter-efficient fine-tuning (PEFT) methods, allowing mobile users to adjust the initial layers of the LLM while edge servers handle the more demanding latter layers. 
Specifically, we formulate a multi-objective optimization problem to minimize the total energy consumption and delay during training.
We also address the common issue of instability in model performance by incorporating stability enhancements into our objective function.
Through novel fractional programming technique, we achieve a stationary point for the formulated problem.
Simulations demonstrate that our method reduces the energy consumption as well as the latency, and increases the reliability of LLMs across various mobile settings.
\end{abstract}



\maketitle
\thispagestyle{empty}
\thispagestyle{fancy}
\pagestyle{fancy}
\lhead{This paper appears in the 2024 International Symposium on Theory, Algorithmic Foundations, and Protocol Design for Mobile Networks and Mobile Computing (MobiHoc).}
\rhead{}
\cfoot{\thepage}
\renewcommand{\headrulewidth}{0.4pt}
\renewcommand{\footrulewidth}{0pt}

\section{Introduction}

The advent of large language models (LLMs) marks a significant milestone in the advancement of artificial intelligence and offers unparalleled capabilities in natural language processing, generation, and understanding.
The desire for ubiquitous access to Artificial intelligence (AI) capabilities has driven a significant trend and demand toward the deployment and even training of these computationally intensive models directly on mobile devices~\cite{cai2023efficient,xu2023llmcad,liu2024mobilellm}. 
Users seek real-time, personalized experiences and decision-making support across a diverse array of applications, from healthcare to customer service, which only such advanced models can provide. 
Additionally, there is a growing emphasis on decentralization in computing to enhance privacy and data security by processing sensitive information locally on the device, rather than transmitting it to distant data centers. 
However, this aspiration faces a tough challenge: the substantial computational resources required by LLMs. These models necessitate sophisticated hardware configurations that far exceed the capabilities of standard mobile devices~\cite{ma2023poster}.

Mobile edge computing (MEC) emerges as a transformative solution to this challenge by bringing computational resources closer to the data source~\cite{shi2016edge}.
MEC enables data processing at the network's edge, utilizing distributed computing resources geographically proximate to where the data originates and is consumed. 
Augmented with LLMs, mobile edge servers can process and understand complex queries locally, minimizing the need for constant communication with centralized cloud infrastructure.
This not only improves response times but also enhances privacy and data security by processing sensitive information closer to its source.

However, we still face challenges in how LLMs can be optimized in real-time mobile applications to achieve the best possible performance. 
The challenges 
involve tailoring these models in resource-constrained environments.
One solution is in-context learning. 
It provides the LLM with a few examples of the desired task within the input prompt, allowing the model to adapt its behavior based on these examples without changing its parameters.
But the effectiveness of in-context learning is constrained by the model's context window size, and it doesn't lead to persistent improvements in the model's capabilities.
Moreover, recent studies have shown that in-context learning may struggle with reliability and hallucination~\cite{huang2024securing}.
Alternatively, many studies propose parameter-efficient fine-tuning (PEFT) methods~\cite{houlsby2019parameter,liu2021p,ding2022delta}.
These methods have been demonstrated to yield state-of-the-art performance in model training and optimization, significantly reducing the computational overhead traditionally associated with such processes.
Inspired by this body of work, we propose a novel scenario for collaboratively training LLMs, harnessing the combined capabilities of mobile users and edge servers. 
In this proposed model, mobile users are responsible for fine-tuning the first several layers of the LLM, capitalizing on the parameter-efficient techniques that require less computational power and are thus more suitable for mobile environments.
At the same time, the edge servers undertake the task of training the remaining layers, leveraging their processing capabilities to manage the more resource-intensive aspects of the training process. 

\begin{table*}[ht]
\caption{A comparative overview of this paper and prior works on MEC.}\vspace{-10pt}
\label{table:related_work}
\centering
\newcolumntype{C}[1]{>{\centering\arraybackslash}m{#1}}
\begin{tabular}{C{2.0cm}|C{2.3cm}|C{2.3cm}|C{6.0cm}}
\hline
\multirow{2}{*}[-1.5ex]{\textbf{Reference}} & \multicolumn{2}{c|}{\textbf{Objective Function}} & \multirow{2}{*}[-1.5ex]{\makecell{\textbf{Optimization technique used to solve} \\ \textbf{the multiplication of variables}}} \\ \cline{2-3}
& \textbf{Energy incorporated} & \textbf{Delay incorporated} & \\ \hline\hline
Mao~\textit{et al.}~\cite{mao2016dynamic}& \scalebox{0.75}{\usym{2613}} & \checkmark &  Exhaustive search-based strategy  \\ \hline
Chen~\textit{et al.}~\cite{chen2018task}   & \scalebox{0.75}{\usym{2613}}& \checkmark& Alternative optimization  \\ \hline
Xu~\textit{et al.}~\cite{xu2023joint}& \scalebox{0.75}{\usym{2613}}& \checkmark&  Deep reinforcement learning-based approach \\ \hline
Zhan~\textit{et al.}~\cite{zhan2020completion} & \checkmark& \scalebox{0.75}{\usym{2613}} & Alternating optimization   \\ \hline
Wang~\textit{et al.}~\cite{wang2017joint}& \checkmark& \scalebox{0.75}{\usym{2613}}&   Lagrange duality method  \\ \hline
\textbf{This paper}  & \checkmark&  \checkmark
 & Novel fractional programming technique \\ \hline
\end{tabular}
\vspace{-7pt}
\end{table*}

Nevertheless, it has been found by many works that the prevalent fine-tuning methods are afflicted with instability concerns~\cite{han2021robust,mosbach2020stability,zhao2021calibrate}. 
The fine-tuning approach, partitioning the computation process between users and edge servers, inherently introduces potential variances in the model's learning dynamics.
When the initial several layers of a large language model are tailored to the idiosyncrasies of local data, these layers may start to generate data representations that are highly specialized or customized to the user's local context. 
Such misalignment can manifest as model instability, where small variations in local data could result in disproportionately large changes in the model's output, reducing its reliability and robustness in real-world applications.
Thus, 
addressing model stability is essential. 
A stable model ensures minor changes in the training process don't lead to large performance discrepancies, making the model robust and reliable across various conditions and data distributions. 
This goal is even more critical in our proposed collaborative training framework, where the training workload is divided between mobile devices and edge servers. 
To solve this problem, we propose to incorporate model stability as a component of our objective function. 
This approach aims to reduce performance variance across training instances, ensuring that the fine-tuning process yields consistently high-quality results, regardless of the minor fluctuations inherent in distributed training environments.
The contributions of this paper are summarized as follows:
\vspace{-2pt}\begin{itemize}[leftmargin=*]
    \item We introduce a collaborative training framework that combines mobile users and edge servers. This framework leverages PEFT methods, allowing mobile users to adjust the initial layers of the LLM while edge servers handle the more demanding latter layers.
    \item We formulate a multi-objective optimization problem that aims to concurrently minimize total energy consumption and user-experienced delay. At the same time, we enhance the stability of LLMs by integrating model stability considerations into our optimization objectives.
    \item To quantify the relationship between the number of the fine-tuned layers and the model stability, we provide the upper bound of the average-replace-one stability through theoretical analysis.
    \item To address the multi-objective optimization problem, we divide the problem into two parts. In the first part, we optimize the offloading decisions and resource allocation through the application of a novel fractional programming technique, which could find a stationary point with local or global optimal guarantee.
    For the second part, the Concave-Convex Procedure (CCCP) is employed to optimize the user-to-edge association problem.
\end{itemize}\vspace{-2pt}
The structure of this paper is laid out as follows: Section \ref{sec-Related-Work} reviews the literature and works related to our study. 
Section \ref{sec-System} outlines the architecture of the MEC-based LLM framework. 
Following that, Section \ref{sec-solution} details the analytical exploration. 
The outcomes of our simulations are presented in Section \ref{sec-Simulation}. 
We conclude the paper with Section \ref{sec-Conclusion}.
\vspace{-6pt}\section{Related Work}\label{sec-Related-Work}
In this section, we review the existing literature related to our work.

\textbf{Resource allocation in mobile edge computing.}
In~\cite{mao2016dynamic}, the authors propose a Lyapunov optimization-based dynamic computation offloading algorithm to optimize the execution delay and the task-dropping cost in MEC system.
When addressing the offloading decision problem, they apply an exhaustive search strategy, assessing the objective values across three discrete options to determine the optimal solution.
Dinh~\textit{et al.}~\cite{dinh2017offloading} propose to minimize the execution latency and the user devices' energy consumption in MEC.
They use an exhaustive search approach and a semidefinite relaxation (SDR)-based approach to optimize the CPU frequency.
However, the exhaustive search approach is not practical in implementation due to its high complexity, and the SDR-based approach has no global or local optimality guarantee.
In~\cite{chen2018task}, Chen~\textit{et al.} optimize the computation resource allocated to each task to minimize the computation and communication delay.
To handle the multiplication of two decision variables (i.e., the computation resource allocation and the offloading decision), they adopt alternative optimization (AO) techniques.
Xu~\textit{et al.}~\cite{xu2023joint} formulate a cooperative resource optimization problem to optimize the offloading decision and resource allocation in vehicular edge computing.
Yet, they decouple the resource allocation variables from the offloading decision variable, and then use a deep reinforcement learning-based approach to solve it.
Zhan~\textit{et al.}~\cite{zhan2020completion} optimize the computation offloading scheduling and resource allocation in unmanned aerial vehicle (UAV)-enabled MEC system.
They propose a two-stage alternating optimization algorithm to optimize the offloading scheduling, resource allocation and time duration alternatively.

In contrast, Wang~\textit{et al.}~\cite{wang2017joint} obtain the optimal solution in a semi-closed form for offloading decisions and resource allocation in MEC with wireless power transfer.
However, their study exclusively focuses on minimizing the total energy consumption without integrating delay considerations into the objective function. 
Consequently, while it facilitates the determination of the optimal CPU frequency, it inherently simplifies the selection process to the minimal processing unit frequency that meets the latency requirements.
Nonetheless, in this paper, we incorporate delay considerations into the objective function, thereby introducing a higher level of complexity to the solution process for resource allocation.
As a result, the optimal solution to our problem cannot be directly ascertained. 
In Table~\ref{table:related_work}, a comparative analysis is presented between this paper and the aforementioned related works.

\textbf{PEFT vs. In-Context Learning (ICL).} 
Recent studies have demonstrated the superiority of PEFT methods over ICL in various scenarios. 
Mosbach~\textit{et al.}~\cite{mosbach-etal-2023-shot} conduct a fair comparison of ICL and fine-tuning approaches across different tasks and model sizes. 
They find that fine-tuning outperforms in-context learning across different performance metrics.
Liu~\textit{et al.}~\cite{NEURIPS2022_0cde695b} also rigorously demonstrate that PEFT surpasses ICL in both accuracy and computational efficiency.

\textbf{The model stability of fine-tuned large language models.}
Extensive efforts have focused on developing algorithms aimed at improving the stability of the fine-tuning process.
Based on the idea of dropout, Lee~\textit{et al.}~\cite{lee2019mixout}  present ``Mixout'' regularization technique to selectively combine the parameters of two pre-trained language models.
This approach effectively regularizes the learning process, improving the stability of the model.
Houlsby~\textit{et al.}~\cite{houlsby2019parameter} propose a transfer method based on a bottleneck adapter architecture.
He~\textit{et al.}~\cite{he2021effectiveness} conduct a comprehensive comparison between two PEFT methods: fine-tuning and adapter-based tuning.
Their works demonstrate that selectively tuning a subset of the parameters from pre-trained models contributes to enhancing the stability of the model. 

For the stability analysis, Fu~\textit{et al.}~\cite{fu2023effectiveness} harmonize the array of PEFT strategies by framing them within the paradigm of sparse fine-tuning models.
They provide a theoretical analysis that highlights sparsity's function as a regulatory mechanism for the original model, effectively imposing a constraint on the upper limit of model stability. 
However, their reliance on pointwise hypothesis stability to evaluate model stability focuses on the sensitivity of individual predictions to changes in the training data.
In contrast, our work employs the average-replace-one stability measure which assesses the model's overall performance variation when a single training instance is replaced. 
In edge computing, we focus more on maintaining high levels of service reliability and efficiency across the entire network, rather than optimizing the outcome for individual predictions. 
Average-replace-one stability aligns with this objective by providing a macroscopic view of model stability.

\vspace{-5pt}\section{System Model}\label{sec-System}\vspace{-2pt}
\begin{figure}[tb]
\captionsetup{justification=justified,singlelinecheck=false}
\includegraphics[width=0.39\textwidth]{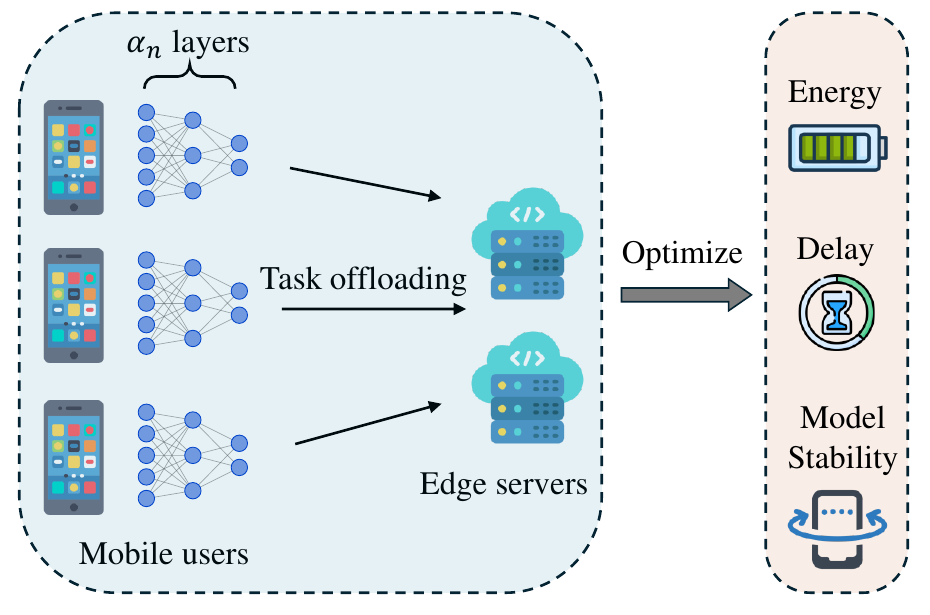} 
\vspace{-10pt}\caption{The proposed system model consists of $N$ mobile users and $M$ edge servers. 
Our optimization problem aims to minimize energy consumption and delay while improving the LLM stability.}
\label{fig_system}\vspace{-18pt}
\end{figure}
In this section, we first present the system model, including local computation model, edge computation model and LLM stability.
After that, we formulate the multi-objective optimization problem.
We consider an MEC system consisting of $N$ users and $M$ edge servers, as described in Figure~\ref{fig_system}.
Assume all the users in the system train LLMs with the same architecture.
Let $\Upsilon$ be the total number of transformer layers in the LLM.
User $n$ fine-tunes the first $\alpha_n$ layers locally, after which the intermediate results are sent to a certain edge server to complete the remaining training process.
Let $d_n$ denote the length of input tokens of user $n$ for training.
For the energy and delay calculation for training LLMs, we follow the setting in~\cite{liu2024resource}.
Let $\psi(d_n)$ be the FLOPs per token required to train one transformer layer, $\psi(d_n) = 72Bd_nh^2+12Bd_n^2h$ where $B$ is the batch size and $h$ is the dimensionality of the hidden states.
\vspace{-8pt}\subsection{Local Computation Model}\vspace{-2pt}
When user $n$ is training one transformer layer locally, the delay for computation can be given by:  
\begin{small}
\vspace{-5pt}\begin{align}
    T_n^{cmp}=\frac{\psi(d_n)}{f_n C_n^{U} D_n^{U}},\\[-17pt]\nonumber
\end{align}
\end{small}
where $f_n$ is the GPU frequency of user $n$, $C_n^{U}$ is the number of cores of the GPU at user $n$ and $D_n^{U}$ is the number of FLOPs per cycle per core of the GPU.
The relationship between the GPU's power consumption and its clock speed is cubic, i.e., $\texttt{power}=\kappa_1f_n^3$.
Here, $\kappa_1$ is the coefficient reflecting the power usage per cubic cycle per second ($[\text{in Watt/(cycle/s)}^3]$), dependent on the specific GPU architecture.
Hence, when training one transformer layer, the energy expenditure for local computations is established as follows:
\begin{small}
\vspace{-13pt}\begin{align}
    E_n^{cmp} = \kappa_1f_n^3 \times T_n^{cmp}=\frac{\kappa_1f_{n}^2\psi(d_n)}{ C_n^{U} D_n^{U}}.\\[-17pt]\nonumber
\end{align}
\end{small}
Upon completing local computations, users transmit the intermediate results to edge servers for further processing.
The association between users and edge servers is represented by  $\chi_{n,m}$ with $\chi_{n,m}=1$ signifying that user $n$ has selected edge server $m$ for further computations,
and $\chi_{n,m}=0$ indicating no such association.
In this context, we adopt Frequency-Division Multiple Access (FDMA) such that communications between users and edge servers are free from mutual interference.
The power used for transmission by user $n$ is denoted as $p_n$. 
Following the principles of the Shannon-Hartley theorem~\cite{cover1999elements}, the transmission rate between user $n$ and edge server $m$ can be formulated as $r_{n,m} = b_{n,m}\log_2(1+\frac{g_{n,m}p_{n}}{\sigma^2b_{n,m}})$,
where $\sigma^2$  represents the power of the noise , $b_{n,m}$  denotes the bandwidth that edge server $m$ assigned to user $n$, $p_n$ is the transmission power of user $n$, and $g_{n,m}$ is the channel gain between user $n$ and edge server $m$.
Let $s(d_n)$ be the size of the intermediate results for user $n$. Therefore, the energy consumption of wireless transmission for user $n$ is:
\begin{small}
\vspace{-8pt}\begin{align}
    E_{n}^{com} =\sum_{m \in \mathcal{M}} \chi_{n,m} \frac{s(d_n)p_n}{r_{n,m}}.\\[-16pt]\nonumber
\end{align}
\end{small}
When user $n$ is training the first $\alpha_n$ layers locally, the computation of both time and energy expenditure is:
\vspace{-3pt}\begin{align}
    Cost_{n}^{u}=\alpha_n \cdot(\omega_t T_n^{cmp} + \omega_e E_n^{cmp})+\omega_eE_n^{com}.\\[-17pt]\nonumber
\end{align}
Here, $\omega_t$ serves as the weighting and normalization factor, reflecting the priority given to minimizing delay, while $\omega_e$ represents the weighting and normalization factor that underscores the importance of reducing energy consumption.
\vspace{-7pt}\subsection{Edge Computation Model}\vspace{-2pt}

When edge server $m$ trains one transformer layer for user $n$, the time taken for the computation can be expressed as follows:
\begin{small}
\vspace{-5pt}\begin{align}
    T^{cmp}_{n,m} = \frac{\psi(d_n)}{f_{n,m} C_m^{E} D_m^{E}},\\[-15pt]\nonumber
\end{align} 
\end{small}
where $f_{n,m}$ denotes the GPU frequency of edge server $m$ assigned to user $n$, and $C_m^{E}$ represents the total core count of the GPU within edge server $m$, and $D_m^{E}$ signifies the computational capability of each core, measured in floating point operations per cycle, for the GPU located at edge server $m$.
Thus, the energy required by edge server $m$ to train one transformer layer for user $n$ can be quantified as follows:
\begin{small}
\vspace{-10pt}\begin{align}
    E_{n,m}^{cmp} =\frac{ \kappa_2 f_{n,m}^2 \psi(d_n)}{C_m^{E}D_m^{E}},\\[-15pt]\nonumber
\end{align} 
\end{small}
where $\kappa_2$ is a coefficient that varies based on the architecture of the chip.
The energy used for downlink transmission from the edge servers to the users is not considered in this calculation, due to the substantially higher power capacities of the edge servers compared to the users.
Furthermore, in comparison to the energy requirements for training the LLM, the energy expended on transmission by the edge servers is considered negligible.

Since there are $\Upsilon$ transformer layers in total, $(\Upsilon - \alpha_n)$ layers are processed at the corresponding edge server.
As a result, the incurred cost for conducting the training tasks for users at edge server $m$ is calculated by integrating both the time delays and energy expenditures into a weighted sum:
\vspace{-5pt}\begin{align}
    Cost_{m}^{E}= \sum_{n \in \mathcal{N}}\chi_{n,m}(\Upsilon-\alpha_n) (\omega_t T_{n,m}^{cmp}+\omega_e E_{n,m}^{cmp}).\\[-23pt]\nonumber
\end{align}

\subsection{LLM Stability}\vspace{-3pt}
In this paper, we use the Average-replace-one Stability (AS) proposed by~\cite{shalev2010learnability} to measure the mode stability. 
AS is a measure of how much an individual prediction is affected by small changes in the training dataset. 
It serves as a crucial metric for ensuring that our fine-tuned language model remains consistent and reliable, despite the variability in local data from user to user.
Next, we give the definition of the average-replace-one stability.
\vspace{-5pt}\begin{defn}[Average-replace-one stability]
Given a loss function $\ell$ and training dataset $\mathcal{S} = \{z_1,\ldots,z_k\}$, an algorithm $\mathcal{A}$ demonstrates the average-replace-one stability (AS) with a bound $\beta$ if the following condition is met:~\mbox{$\forall i \in \{1, \ldots, k\}$,}
\vspace{-3pt}\begin{align}
\mathbb{E}_{\mathcal{S}} \left[ \left| \ell(\mathcal{A}(\mathcal{S}), z_i) - \ell(\mathcal{A}(\mathcal{S}^{i}), z_i) \right| \right] \leq \beta,\\[-18pt]\nonumber
\end{align}
where $\mathcal{A}(\mathcal{S})$ denotes the model obtained after the algorithm $\mathcal{A}$ has been trained on the dataset $\mathcal{S}$, 
and $\ell(\mathcal{A}(\mathcal{S}), z_i)$ is the loss function evaluated at a particular data point $z_i$ using the model given by $\mathcal{A}(\mathcal{S})$.
$\mathcal{S}^{i}$ represents the training dataset with the $i$-th sample replaced with $z_i^\prime$, i.e.,~\mbox{$\mathcal{S}^{i} = \{z_1,\ldots,z_{i-1},z_i^\prime,\ldots,z_k\}$.}
\end{defn}\vspace{-3pt}
This definition implies that for every individual element $z_i$ in a dataset of size $k$, the expected disparity in the loss computed by algorithm $\mathcal{A}$ when trained with the complete dataset versus the dataset lacking that specific sample is bounded by $\beta$.

\vspace{-6pt}\subsection{Problem Formulation}\vspace{-3pt}
With the computation and communication model above, we then formulate the joint optimization problem that aims to minimize the system's cost while minimizing the Average-replace-one Stability (AS) of the LLMs, by optimizing the following variables: the number of transformer layers that execute locally: \mbox{$\boldsymbol{\alpha}:=[\alpha_n|_{n \in \mathcal{N}}]$}, the user-to-edge server association: $\boldsymbol{\chi}:=[\chi_{n,m}|_{n \in \mathcal{N}, m \in \mathcal{M}}]$, the transmission power of the users: $\boldsymbol{p}:=[p_n|_{n \in \mathcal{N}}]$, the bandwidth allocation: \mbox{$\boldsymbol{b}:=[b_{n,m}|_{n \in \mathcal{N}, m \in \mathcal{M}}]$}, the users' GPU frequency: $\boldsymbol{f^U}:=[f_{n}|_{n \in \mathcal{N}}]$ and the edge servers' GPU frequency allocation: $\boldsymbol{f^E}:=[f_{n,m}|_{n \in \mathcal{N}, m \in \mathcal{M}}]$.
Similar to delay and energy, we also give a weighting and normalization parameter $\omega_s$ to the AS.
The joint optimization problem is formulated as follows:
\vspace{-5pt}\begin{small}
\begin{subequations}\label{P1}
\begin{align}  \text{Problem}~\mathbb{P}_1:&\!\!\min_{\boldsymbol{\alpha,\chi,p,b,f^U,f^E}} \sum_{n \in \mathcal{N}} Cost_{n}^{u} + \sum_{m \in \mathcal{M}} Cost_{m}^{E}+\omega_s AS\tag{\ref{P1}},\nonumber\\
    &\text{~~~s.t.~}\alpha_n \in \{1,2,\ldots,\Upsilon\},\forall n \in \mathcal{N},\label{P1_C_alpha}\\
    &\phantom{~~~s.t.~}\chi_{n,m}\in \{0,1\},\forall n \in \mathcal{N}, m \in \mathcal{M},\label{P1_C_chi_1}\\
    &\phantom{~~~s.t.}\sum_{m \in \mathcal{M}}\chi_{n,m}=1, \forall n \in \mathcal{N},\label{P1_C_chi_2}\\
    &\phantom{~~~s.t.~}p_n \leq p_n^{max},\forall n \in \mathcal{N},\label{P1_C_pn}\\
    &\phantom{~~~s.t.}\sum_{n \in \mathcal{N}} \chi_{n,m} b_{n,m}=b_m^{max},\forall m \in \mathcal{M},\label{P1_C_bn_2}\\
    &\phantom{~~~s.t.~} f_n \leq f_n^{max},\forall n \in \mathcal{N},\label{P1_C_f_n}\\
    &\phantom{~~~s.t.}\sum_{n \in \mathcal{N}}\chi_{n,m} f_{n,m}=f_m^{max},\forall m \in \mathcal{M}.\label{P1_C_fnm_2}
\end{align}
\end{subequations}    
\end{small}
Given the inherent challenges in quantifying the average-replace-one stability, we commence by presenting the following theorem to facilitate addressing the optimization problem.
We assume the loss function $\ell(\cdot)$ is $L$-Lipschitz and strong convex. 
These two assumptions are widely employed in the analysis of the behavior of neural networks~\cite{shalev2014understanding,pmlr-v178-schliserman22a}.
\vspace{-5pt}\begin{thm}\label{theorem_AS}
If a user fine-tunes a proportion $\alpha$ of the parameters, the expectation of the loss has an AS bounded by $\frac{2L^2}{k(1-\alpha)}$. I.e., $\forall i \in \{1, \ldots, k\},$
\vspace{-8pt}\begin{align}
     \mathbb{E}_{\mathcal{S}} \left[ \left| \ell(\mathcal{A}(\mathcal{S}), z_i) -\ell(\mathcal{A}(\mathcal{S}^{i}), z_i) \right| \right] \leq\frac{2L^2}{k(1-\alpha)}.
\end{align}
\end{thm}
\vspace{-13pt}\begin{proof}
The proof can be found in Appendix \ref{Appendix.A}.
\end{proof}\vspace{-8pt}
Theorem \ref{theorem_AS} provides a quantifiable measure of model stability and bridges the concept of ``model stability'' with a measurable quantity. 
Since the ``model stability'' term is not quantitative in problem $\mathbb{P}_1$,
we re-formulate problem $\mathbb{P}_1$ into the following $\mathbb{P}_2$ by replacing the sum of AS with the sum of the AS's upper bound of all the users:
\begin{small}
\begin{subequations}\label{P2}
\vspace{-6pt}\begin{align}
\text{Problem}~\mathbb{P}_2:&\!\!\!\!\!\!\min_{\boldsymbol{\alpha,\chi,p,b,f^U,f^E}}\!\sum_{n \in \mathcal{N}}\!\! Cost_{n}^{u} \!+\!\! \!\sum_{m \in \mathcal{M}} \!\!\!Cost_{m}^{E}\!+\omega_s \!\!\sum_{n \in \mathcal{N}}\frac{2L^2}{k_n(1\!-\!\frac{\alpha_n}{\Upsilon})}\tag{\ref{P2}},\!\\[-1pt]
&\text{s.t.~(\ref{P1_C_alpha})~--~(\ref{P1_C_fnm_2})}.\nonumber\\[-17pt]\nonumber
\end{align}
\end{subequations}
\end{small}
While the optimal solutions to problem $\mathbb{P}_1$ and $\mathbb{P}_2$ may not be strictly equivalent in a mathematical sense, $\mathbb{P}_2$ serves as a practical approximation of $\mathbb{P}_1$. 
By using the upper bound from Theorem \ref{theorem_AS}, we are optimizing for the worst-case scenario of model instability.
Problem $\mathbb{P}_2$ falls into the category of Mixed Integer Nonlinear Programming (MINLP) problem.
This classification arises due to the inclusion of both integer-valued decision variables and nonlinear terms involving products of variables, a combination that inherently induces non-convexity in the problem space.
The non-convex nature of this problem makes it especially challenging to solve because it cannot be addressed using standard optimization methods, which typically rely on the problem being convex.
In order to tackle the non-convex problem, we optimize $\boldsymbol{\alpha,p,b,f^U,f^E}$ and $\boldsymbol{\chi}$ iteratively.
Specifically, in the first step, we fix $\boldsymbol{\chi}$  and utilize a novel fractional programming technique motivated by Zhao \textit{et al.}~\cite{zhao2023human} to optimize $\boldsymbol{\alpha,p,b,f^U,f^E}$ by transforming the non-convex problem into a series of parametric convex problems.
In the second step, given $\boldsymbol{\alpha,p,b,f^U,f^E}$, the method of CCCP is adopted to facilitate the solution to $\boldsymbol{\chi}$ by solving a sequence of convex problems.
\vspace{-5pt}\section{Proposed Algorithm}\label{sec-solution}
In this section, we provide a detailed solution to the optimization problem.
\vspace{-5pt}\subsection{Optimizing $\boldsymbol{\alpha,p,b,f^U,f^E}$ given $\boldsymbol{\chi}$}
The discrete variable $\alpha_n$ is difficult to handle.
Thus, we first relax $\alpha_n$ to continuous variables, which will be rounded back to the nearest integer later.
For problem $\mathbb{P}_2$, to optimize $\boldsymbol{\alpha,p,b,f^U,f^E}$ given $\boldsymbol{\chi}$ means to solve the following optimization problem:
\begin{small}
\begin{subequations}\label{P3_chi}
\vspace{-3pt}\begin{align}
&\text{Problem}~\mathbb{P}_3(\boldsymbol{\chi}):\min_{\boldsymbol{\alpha,p,b,f^U,f^E}} H(\boldsymbol{\alpha,p,b,f^U,f^E})=\nonumber\\[-3pt]&\sum_{n \in \mathcal{N}} Cost_{n}^{u} \!+\!\! \sum_{m \in \mathcal{M}} \!Cost_{m}^{E}\!+\omega_s \!\sum_{n \in \mathcal{N}}\frac{2L^2}{k_n\cdot(1-\frac{\alpha_n}{\Upsilon})}\tag{\ref{P3_chi}},\\
&\text{s.t.~}1\leq \alpha
_n \leq \Upsilon, \forall n \in \mathcal{N},\label{P3_C_alpha}\\
&\phantom{s.t.~}\text{(\ref{P1_C_pn})--(\ref{P1_C_fnm_2})}.\nonumber\\[-17pt]\nonumber
\end{align}
\end{subequations}
\end{small}
Problem $\mathbb{P}_3$ involves fraction term and multiplication terms, which makes it difficult to solve using standard optimization algorithms.
Motivated by the novel fractional programming technique proposed in~\cite{zhao2023human}, we next transform problem $\mathbb{P}_3$ into a series of $\mathbb{P}_4$:
\begin{small}
\begin{subequations}\label{P4}
\vspace{-5pt}\begin{align}
&\text{Problem}~\mathbb{P}_4(\boldsymbol{\chi},\boldsymbol{z},\boldsymbol{\nu},\boldsymbol{q}):\!\!\!\!\min_{\boldsymbol{\alpha,p,b,f^U,f^E}} K(\boldsymbol{\alpha,p,b,f^U,f^E,z,\nu,q})=\nonumber\\[-1pt]
&\sum_{n \in \mathcal{N}}\Big(\alpha_n^2z_{n}+\frac{(\omega_t\frac{\psi(d_n)}{f_nC_n^{U}D_n^{U}}+\omega_e\frac{\kappa_1 f_n^2 \psi(d_n)}{C_n^{U}D_n^{U}})^2}{4z_{n}}\Big)+\nonumber\\[-4pt]
&\quad\!\!\!\omega_e \sum_{n \in \mathcal{N}} \sum_{m \in \mathcal{M}} \chi_{n,m}\Big((p_n d_n)^2\nu_{n,m} + \frac{1}{4r^2_{n,m}\nu_{n,m}}\Big)+\nonumber\\[-4pt]
&\sum_{n \in \mathcal{N}} \!\sum_{m \in \mathcal{M}}\!\!\chi_{n,m}\Big( (\Upsilon\!-\!\alpha_n)^2 q_{n,m}\!+\!\frac{\Big(\omega_t \frac{\psi(d_n)}{f_{n,m}C^{E}_m D^{E}_m} \!+\! \omega_e \frac{\kappa_2 f_{n,m}^2 \psi(d_n)}{C^{E}_m D^{E}_m}\Big)^2}{4q_{n,m}}\Big)\nonumber\\[-2pt]
& + \omega_s \sum_{n \in \mathcal{N}} \frac{2L^2}{k_n\cdot(1-\frac{\alpha_n}{\Upsilon})},\tag{\ref{P4}}\\[-2pt]
&\text{s.t.~(\ref{P3_C_alpha}),~(\ref{P1_C_pn})--(\ref{P1_C_fnm_2})},\nonumber\\[-20pt]\nonumber
\end{align}
\end{subequations}
\end{small}
where the auxiliary variables $\boldsymbol{z}:=[z_1,z_2,\ldots,z_n]$ with $z_n > 0$, $\boldsymbol{\nu}:=[\nu_{1,1},\nu_{1,2},\ldots,\nu_{1,m},\ldots,\nu_{n,m}]$ with $\nu_{n,m} > 0$ and $\boldsymbol{q}:=[q_{1,1},q_{1,2},\ldots,$ $q_{1,m},\ldots,q_{n,m}]$  with $q_{n,m} > 0$.

Problem $\mathbb{P}_3$ involves non-convex terms and is difficult to handle.
Therefore, we formulate the above problem $\mathbb{P}_4$.
Next, we introduce an AO algorithm for problem $\mathbb{P}_4$.
After that, we propose Proposition \ref{proposition1} to explain how we can tackle problem $\mathbb{P}_3$ through a series of convex problem $\mathbb{P}_4$ instances.

First, we introduce the AO algorithm for problem $\mathbb{P}_4$. Overall, we alternatively optimize $\boldsymbol{\alpha,p,b,f^U,f^E}$ and $\boldsymbol{z},\boldsymbol{\nu},\boldsymbol{q}$.
Specifically, we begin with an initial feasible $[\boldsymbol{\alpha}^{(0)},\boldsymbol{p}^{(0)},$ $\boldsymbol{b}^{(0)},\boldsymbol{f^U}^{(0)},$ $\boldsymbol{f^E}^{(0)}]$.
Next, we denote $A(f_n)$ and $B(f_{n,m})$ as:
\begin{small}
\vspace{-5pt}\begin{align}
A(f_n)&=\omega_t\frac{\psi(d_n)}{f_nC_n^{U}D_n^{U}}+\omega_e\frac{\kappa_1 f_n^2 \psi(d_n)}{C_n^{U}D_n^{U}},\\[-1pt]
B(f_{n,m})&=\omega_t \frac{\psi(d_n)}{f_{n,m}C^{E}_m D^{E}_m} \!+\! \omega_e \frac{\kappa_2 f_{n,m}^2 \psi(d_n)}{C^{E}_m D^{E}_m}.\\[-15pt]\nonumber
\end{align}
\end{small}
We assign $z_n^{(0)}$ to be $\frac{A(f_n^{(0)})}{2\alpha_n^{(0)}}$, 
which is the optimal value of $z_n$ when optimizing $\alpha_n^2 z_n + \frac{A^2(f_n)}{4 z_n}$ with respect to $z_n$, 
while keeping $\alpha_n,f_n$ fixed at $\alpha_n^{(0)},f_n^{(0)}$;
we assign $\nu_{n,m}^{(0)}$ to be $\frac{1}{2p_n^{(0)} d_n^{(0)} r_{n,m}^{(0)}}$, 
which is the optimal value of $\nu_{n,m}$ when optimizing $(p_n d_n)^2\nu_{n,m} + \frac{1}{4r^2_{n,m}\nu_{n,m}}$ with respect to $\nu_{n,m}$, 
while keeping $p_n,d_n$ fixed at $p_n^{(0)},d_n^{(0)}$;
we assign $q_{n,m}^{(0)}$ to be $\frac{B(f_{n,m}^{(0)})}{2(\Upsilon-\alpha_n^{(0)})}$, 
which is the optimal value of $q_{n,m}$ when optimizing $(\Upsilon-\alpha_n)^2 q_{n,m} + \frac{B^2(f_{n,m})}{4 q_{n,m}}$ with respect to $q_{n,m}$, 
while keeping $\alpha_n,f_{n,m}$ fixed at $\alpha_n^{(0)},f_{n,m}^{(0)}$.

After that, we solve problem $\mathbb{P}_4(\boldsymbol{\chi},\boldsymbol{z}^{(0)},\boldsymbol{\nu}^{(0)},\boldsymbol{q}^{(0)})$, 
from which we derive the solution $[\boldsymbol{\alpha}^{(1)},\boldsymbol{p}^{(1)},\boldsymbol{b}^{(1)},\boldsymbol{f^U}^{(1)},\boldsymbol{f^E}^{(1)}]$, and subsequently update $[\boldsymbol{z}^{(1)},\boldsymbol{\nu}^{(1)},\boldsymbol{q}^{(1)}]$.
 This procedure is repeated in an iterative fashion: during the $(t+1)$-th iteration,
 we set $z_n^{(t)},\nu_{n,m}^{(t)}$ and $q_{n,m}^{(t)}$ as $\frac{A(f_n^{(t)})}{2\alpha_n^{(t)}}, \frac{1}{2p_n^{(t)} d_n^{(t)} r_{n,m}^{(t)}}$ and $\frac{B(f_{n,m}^{(t)})}{2(\Upsilon-\alpha_n^{(t)})}$, and then solve $\mathbb{P}_4(\boldsymbol{\chi},\boldsymbol{z}^{(t)},\boldsymbol{\nu}^{(t)},\boldsymbol{q}^{(t)})$, to obtain $[\boldsymbol{\alpha}^{(t+1)},\boldsymbol{p}^{(t+1)},\boldsymbol{b}^{(t+1)},\boldsymbol{f^U}^{(t+1)},$\phantom{s} 
 $\boldsymbol{f^E}^{(t+1)}]$.
The above AO process converges when the difference between the objective function of problem $\mathbb{P}_4(\boldsymbol{\chi},\boldsymbol{z}^{(t-1)},\!\boldsymbol{\nu}^{(t-1)}\!,\boldsymbol{q}^{(t-1)})$ and problem $\mathbb{P}_4(\boldsymbol{\chi},\boldsymbol{z}^{(t)},\boldsymbol{\nu}^{(t)},\boldsymbol{q}^{(t)})$ falls below a predefined small error tolerance.
Then, we propose the following proposition to explain how we solve $\mathbb{P}_3$ through the AO process for $\mathbb{P}_4$.

\begin{prop}\label{proposition1}
We can derive a stationary point for \mbox{problem $\mathbb{P}_3$} by applying the AO process outlined above for \mbox{problem $\mathbb{P}_4$} until convergence.
\end{prop}
\vspace{-12pt}\begin{proof}
Denote ``$\boldsymbol{\alpha,p,b,f^U,f^E}$" as ``$\bigstar$" and ``$\boldsymbol{z,\nu,q}$" as ``$\blacklozenge$".
In the first step in the AO process, we optimize $\blacklozenge$ while keeping $\bigstar$ fixed, i.e., letting $z_n^{\#}=\frac{A(f_n)}{2\alpha_n}$, $\nu_{n,m}^{\#}=\frac{1}{2p_n d_n r_{n,m}}$,  $q_{n,m}^{\#}=\frac{B(f_{n,m})}{2(\Upsilon-\alpha_n)}$.
When we substitute back $z_n^{\#}, \nu_{n,m}^{\#}, q_{n,m}^{\#}$ to $K(\bigstar,\blacklozenge)$, $K(\bigstar,\blacklozenge)$ will become $H(\bigstar)$, i.e.,
\vspace{-5pt}\begin{align}
K(\bigstar,\blacklozenge)|_{z_n=z_n^{\#},\nu_{n,m}=\nu_{n,m}^{\#},q_{n,m}=q_{n,m}^{\#}}=H(\bigstar).\label{zhao_84}\\[-17pt]\nonumber
\end{align}
Next, we investigate the partial derivative of $K(\bigstar,\blacklozenge)$ w.r.t $\bigstar$:
\begin{small}
\vspace{-5pt}\begin{align}
    &\frac{\partial K(\bigstar,\blacklozenge)}{\partial\alpha_n} =\dfrac{2L^2{\omega}_s}{k_n{\Upsilon}\!\cdot\!\left(1\!-\!\frac{{\alpha}_n}{{\Upsilon}}\right)^2}\!+\!2z_n{\alpha}_n\!-\!2\!\!\sum_{m \in \mathcal{M}}\!\!\chi_{n,m}q_{n,m}\!\cdot\!\left(\Upsilon\!-\!\alpha_n\right),\label{partial_K_alpha}\\[-5pt]
    &\frac{\partial K(\bigstar,\blacklozenge)}{\partial p_n} = \sum_{m \in \mathcal{M}}\chi_{n,m}{\omega}_e\cdot \nonumber\\[-8pt]& \Big(2d_n^2{\nu}_{n,m}p_n\!-\!\dfrac{\ln^2\left(2\right)\,g_{n,m}}{2b_{n,m}^3{\nu}_{n,m}{\sigma}^2\!\cdot\!\left(\frac{g_{n,m}p_n}{b_{n,m}{\sigma}^2}\!+\!1\right)\ln^3\left(\frac{g_{n,m}p_n}{b_{n,m}{\sigma}^2}\!+\!1\right)}\Big)\\[-2pt]
    &\frac{\partial K(\bigstar,\blacklozenge)}{\partial b_{n,m}} = \frac{\chi_{n,m}{\omega}_e \ln^2\left(2\right)}{2{\nu}_{n,m} \ln^2\left(\frac{g_{n,m}p_n}{{\sigma}^2b_{n,m}}\!+\!1\right) b_{n,m}^3}\cdot \nonumber \\[-6pt] 
    & \left(\dfrac{g_{n,m}p_n}{{\sigma}^2\ln\left(\frac{g_{n,m}p_n}{{\sigma}^2b_{n,m}}\!+\!1\right)\!\cdot\!\left(\frac{g_{n,m}p_n}{{\sigma}^2b_{n,m}}\!+\!1\right)b_{n,m}}-1\right), \\
     &\frac{\partial K(\bigstar,\blacklozenge)}{\partial f_n} = \dfrac{\psi^2(d_n)\cdot\left({\kappa}_1{\omega}_ef_n^3+{\omega}_t\right)\cdot\left(2{\kappa}_1{\omega}_ef_n^3-{\omega}_t\right)}{2z_nf_n^3\cdot(C_n^{U}D_n^{U})^2},\\
      &\frac{\partial K(\bigstar,\blacklozenge)}{\partial f_{n,m}} =\dfrac{\chi_{n,m}\psi^2(d_n)\cdot\left({\kappa}_2{\omega}_ef_{n,m}^3+{\omega}_t\right)\cdot\left(2{\kappa}_2{\omega}_ef_{n,m}^3-{\omega}_t\right)}{2q_{n,m}f_{n,m}^3\cdot(C_m^{E}D_m^{E})^2}.\label{partial_K_f}
\end{align}
\end{small}
From (\ref{partial_K_alpha}) to (\ref{partial_K_f}), it can be found that
\begin{small}
\vspace{-3pt}\begin{align}
    &\Big(\frac{\partial K(\bigstar,\blacklozenge)}{\partial\alpha_n} \Big)|_{z_n=\frac{A(f_n)}{2\alpha_n},q_{n,m}=\frac{B(f_{n,m})}{2(\Upsilon-\alpha_n)}} = \nonumber \\[-3pt]& \dfrac{2L^2{\omega}_s}{k_n{\Upsilon}\cdot\left(1-\frac{{\alpha}_n}{{\Upsilon}}\right)^2}+A(f_n)-\sum_{m \in \mathcal{M}} \chi_{n,m}\cdot B(f_{n,m}),\label{partial_K_alpha_z}\\[-3pt]
    &\Big(\frac{\partial K(\bigstar,\blacklozenge)}{\partial p_n} \Big)|_{\nu_{n,m}=\frac{1}{2p_n d_n r_{n,m}}}= \sum_{m \in \mathcal{M}}\chi_{n,m}{\omega}_e\cdot \nonumber\\[-5pt]&\left(\frac{d_n}{b_{n,m}\log_2(1\!+\!\frac{g_{n,m}p_{n}}{\sigma^2b_{n,m}})}\!-\!\frac{\ln(2)g_{n,m} p_n d_n}{b_{n,m} (g_{n,m}p_n\!+\!b_{n,m} \sigma^2)\ln^2(\frac{g_{n,m}p_n}{b_{n,m}{\sigma}^2}\!+\!1)}\right),\\[-5pt]
    &\Big(\frac{\partial K(\bigstar,\blacklozenge)}{\partial b_{n,m}} \Big)|_{\nu_{n,m}=\frac{1}{2p_n d_n r_{n,m}}}= \frac{\chi_{n,m} \omega_e \ln(2) p_n d_n }{b_{n,m}^2\ln(\frac{g_{n,m}p_n}{b_{n,m}{\sigma}^2}\!+\!1)}\cdot \nonumber \\[-4pt]&\left(\dfrac{g_{n,m}p_n}{{\sigma}^2\ln\left(\frac{g_{n,m}p_n}{{\sigma}^2b_{n,m}}\!+\!1\right)\!\cdot\!\left(\frac{g_{n,m}p_n}{{\sigma}^2b_{n,m}}\!+\!1\right)b_{n,m}}-1\right),\\
    &\Big(\frac{\partial K(\bigstar,\blacklozenge)}{\partial f_n} \Big)|_{z_n=\frac{A(f_n)}{2\alpha_n}}= \frac{\alpha_n \psi(d_n)\cdot(2\kappa_1 \omega_e f_n^3- \omega_t)}{f_n^2 C_n^V D_n^V},\\
    &\Big(\frac{\partial K(\bigstar,\blacklozenge)}{\partial f_{n,m}} \Big)|_{q_{n,m}\!=\!\frac{B(f_{n,m})}{2(\Upsilon-\alpha_n)}}\!\!=\! \frac{(\Upsilon\!-\!\alpha_n) \chi_{n,m} \psi(d_n) \!\!\cdot\!\! (2\kappa_2 \omega_e f_{n,m}^3\!-\!\omega_t)}{f_{n,m}^2 C_m^R D_m^R}\!.\!\label{partial_K_f_q}
\end{align}
\end{small}
Besides, the partial derivative of $H(\bigstar)$ is given by:
\begin{small}
\begin{align}
&\frac{\partial H(\bigstar)}{\partial \alpha_n} = \omega_t \frac{\psi(d_n)}{f_n C_n^{U} D_n^{U}} + \omega_e \frac{\kappa_1f_{n}^2\psi(d_n)}{ C_n^{U} D_n^{U}} + \dfrac{2L^2{\omega}_s}{k_n{\Upsilon}\cdot\left(1-\frac{{\alpha}_n}{{\Upsilon}}\right)^2}- \nonumber \\ &\sum_{m \in \mathcal{M}} \chi_{n,m} (\omega_t \frac{\psi(d_n)}{f_{n,m} C_m^{E} D_m^{E}}+\omega_e \frac{ \kappa_2 f_{n,m}^2 \psi(d_n)}{C_m^{E}D_m^{E}}),\label{partial_H_alpha}\\
&\frac{\partial H(\bigstar)}{\partial p_n} =\sum_{m \in \mathcal{M}} \ln\left(2\right) \omega_e\chi_{n,m} d_n\cdot \nonumber\\&\dfrac{ \left(g_{n,m}p_n+b_{n,m}{\sigma}^2\right)\ln(\frac{g_{n,m}p_n}{b_{n,m}{\sigma}^2}+1)-g_{n,m}p_n}{b_{n,m}\cdot\left(g_{n,m}p_n+b_{n,m}{\sigma}^2\right)\ln^2(\frac{g_{n,m}p_n}{b_{n,m}{\sigma}^2}+1)},  \\[-4pt]
&\frac{\partial H(\bigstar)}{\partial b_{n,m}} \!= \dfrac{\omega_e\ln\left(2\right)\,\chi_{n,m} d_ng_{n,m}p_n^2}{{\sigma}^2\ln^2\left(\frac{g_{n,m}p_n}{{\sigma}^2 b_{n,m}}\!+\!1\right)\!\left(\frac{g_{n,m}p_n}{{\sigma}^2b_{n,m}}\!+\!1\right)b_{n,m}^3}\!-\!\dfrac{\omega_e\ln\left(2\right)\,\chi_{n,m}d_np_n}{\ln\left(\frac{g_{n,m}p_n}{{\sigma}^2b_{n,m}}\!+\!1\right)b_{n,m}^2}\!,\\[-6pt]
&\frac{\partial H(\bigstar)}{\partial f_n} = -\frac{\alpha_n \omega_t \psi(d_n)}{C_n^V D_n^V f_n^2} + \frac{2\alpha_n \omega_e \kappa_1 f_n \psi(d_n)}{C_n^V D_n^V}, \\[-3pt]
&\frac{\partial H(\bigstar)}{\partial f_{n,m}} = \chi_{n,m} (\Upsilon-\alpha_n) \Big(\frac{2\omega_e \kappa_2 f_{n,m}\psi(d_n)}{C_m^R D_m^R}-\frac{\omega_t \psi(d_n)}{C_m^R D_m^R f_{n,m}^2}\Big).\label{partial_H_f}\\[-17pt]\nonumber
\end{align}
\end{small}
From (\ref{partial_K_alpha_z})--(\ref{partial_K_f_q}) and (\ref{partial_H_alpha})--(\ref{partial_H_f}), it can be observed that:
\begin{small}
\begin{subequations}
\begin{align}
    &\frac{\partial H(\bigstar)}{\partial \alpha_n} = \Big(\frac{\partial K(\bigstar,\blacklozenge)}{\partial\alpha_n} \Big)|_{z_n=\frac{A(f_n)}{2\alpha_n},q_{n,m}=\frac{B(f_{n,m})}{2(\Upsilon-\alpha_n)}},\label{zhao_82_1}\\
     &\frac{\partial H(\bigstar)}{\partial p_n} = \Big(\frac{\partial K(\bigstar,\blacklozenge)}{\partial p_n} \Big)|_{\nu_{n,m}=\frac{1}{2p_n d_n r_{n,m}}},\\
     &\frac{\partial H(\bigstar)}{\partial b_{n,m}} = \Big(\frac{\partial K(\bigstar,\blacklozenge)}{\partial b_{n,m}} \Big)|_{\nu_{n,m}=\frac{1}{2p_n d_n r_{n,m}}},\\
     &\frac{\partial H(\bigstar)}{\partial f_n}  = \Big(\frac{\partial K(\bigstar,\blacklozenge)}{\partial f_n} \Big)|_{z_n=\frac{A(f_n)}{2\alpha_n}},\\
     &\frac{\partial H(\bigstar)}{\partial f_{n,m}}  = \Big(\frac{\partial K(\bigstar,\blacklozenge)}{\partial f_{n,m}} \Big)|_{q_{n,m}=\frac{B(f_{n,m})}{2(\Upsilon-\alpha_n)}}.\label{zhao_82_2}
\end{align}
\end{subequations}
\end{small}
The process of AO in minimizing $K(\bigstar,\blacklozenge)$ is non-increasing.
To elaborate, it's observed that $K(\bigstar^{(i)},\blacklozenge^{(i)}) \leq K(\bigstar^{(i-1)},\blacklozenge^{(i)}) \leq K(\bigstar^{(i-1)},\blacklozenge^{(i-1)})$.
Thus, $K(\bigstar^{(i)},\blacklozenge^{(i)})$ tends towards convergence as the iteration count $i$ increases infinitely.
Assume after the AO process, $\bigstar$ and $\blacklozenge$ converges to $(\bigstar^*,\blacklozenge^*)$.
It indicates:
\textcircled{1} $\blacklozenge^*$ is the optimial value of $\blacklozenge$ if we fix $\bigstar$ as $\bigstar^*$ and minimize $K(\bigstar^*,\blacklozenge)$, and
\textcircled{2} $\bigstar^*$ is the optimial value of $\bigstar$ if we fix $\blacklozenge$ as $\blacklozenge^*$ and minimize $K(\bigstar,\blacklozenge^*)$.

From result \textcircled{2}, we know that $\bigstar^*$ satisfies the Karush--Kuhn--Tucker (KKT) conditions of problem $\mathbb{P}_4$.
With $\boldsymbol{\gamma}, \boldsymbol{\mu}$ denoting the Lagrange multipliers for the inequality constraints and equality constraints, respectively, the Lagrangian function of problem $\mathbb{P}_4$ is given by:
\begin{small}
\begin{align}
    &L_{\mathbb{P}_4}(\bigstar,\blacklozenge,\boldsymbol{\gamma}, \boldsymbol{\mu})= K(\bigstar,\blacklozenge)+
    \!\sum_{n \in \mathcal{N}}\! \Big(\gamma_{1,n}(1-\alpha_n)+\nonumber\\&\gamma_{2,n}(\alpha_n-\!\Upsilon) + \gamma_{3,n} (p_n-p_n^{max}) + \gamma_{4,n}(f_n-f_n^{max})\Big)+\nonumber\\
    &\sum_{m \in \mathcal{M}}\!\Big(\mu_{1,m} ( \!\sum_{n \in \mathcal{N}} \!\chi_{n,m} b_{n,m}\!-\!b_m^{max})\!+\!\mu_{2,m}(\sum_{n \in \mathcal{N}}\chi_{n,m} f_{n,m}\!-\!f_m^{max})\Big).
\end{align}
\end{small}
The KKT conditions of problem $\mathbb{P}_4$ is given by:
\\[-5pt]
\begin{numcases}{}
\text{Stationarity:}&  \nonumber\\
$$\frac{\partial L_{\mathbb{P}_4}}{\partial \boldsymbol{\alpha}^*}=\frac{\partial L_{\mathbb{P}_4}}{\partial \boldsymbol{p}^*}=\frac{\partial L_{\mathbb{P}_4}}{\partial \boldsymbol{b}^*}=\frac{\partial L_{\mathbb{P}_4}}{\partial \boldsymbol{f^U}^*}=\frac{\partial L_{\mathbb{P}_4}}{\partial \boldsymbol{f^E}^*}=0$$,  \label{p4_kkt_1} & \\
\text{Primal feasibility:}&  \nonumber \\
$$ \bigstar^*\text{~satisfy (\ref{P3_C_alpha}),~(\ref{P1_C_pn})~--~(\ref{P1_C_fnm_2})}$$,\label{p4_kkt_2}&\\
\text{Dual feasibility:}&  \nonumber \\
$$\gamma_{1,n}, \gamma_{2,n}, \gamma_{3,n}, \gamma_{4,n} \geq 0, \forall n \in \mathcal{N}$$,\label{p4_kkt_3} &\\
\text{Complementary slackness:} &  \nonumber\\
$$\gamma_{1,n}(1-\alpha_n^*)=0,  \gamma_{3,n} (p_n^*-p_n^{max}) =0$$,\nonumber & \\ 
$$\gamma_{2,n}(\alpha_n^*-\!\Upsilon) =0, \gamma_{4,n}(f_n^*-f_n^{max})=0, \forall n \in \mathcal{N}$$.&\label{p4_kkt_4} 
\end{numcases}
\\[+5pt]
We rewrite $L_{\mathbb{P}_4}$ as $L_{\mathbb{P}_4}=K(\bigstar,\blacklozenge) + \mathcal{Q}(\bigstar,\boldsymbol{\gamma}, \boldsymbol{\mu})$.
Then, (\ref{p4_kkt_1}) can be rewritten as:
\begin{small}
\begin{subequations}    
\begin{align}
    \frac{\partial L_{\mathbb{P}_4}}{\partial \boldsymbol{\alpha}} = \frac{\partial K(\bigstar^*,\blacklozenge^*)}{\partial \boldsymbol{\alpha}} + \frac{\partial \mathcal{Q}(\bigstar^*,\boldsymbol{\gamma}, \boldsymbol{\mu})}{\partial \boldsymbol{\alpha}}=0,\label{p4_L_K_Q_1}\\ 
    \frac{\partial L_{\mathbb{P}_4}}{\partial \boldsymbol{p}} = \frac{\partial K(\bigstar^*,\blacklozenge^*)}{\partial \boldsymbol{p}} + \frac{\partial \mathcal{Q}(\bigstar^*,\boldsymbol{\gamma}, \boldsymbol{\mu})}{\partial \boldsymbol{p}}=0,\\
    \frac{\partial L_{\mathbb{P}_4}}{\partial \boldsymbol{b}} = \frac{\partial K(\bigstar^*,\blacklozenge^*)}{\partial \boldsymbol{b}} + \frac{\partial \mathcal{Q}(\bigstar^*,\boldsymbol{\gamma}, \boldsymbol{\mu})}{\partial \boldsymbol{b}}=0,\\
    \frac{\partial L_{\mathbb{P}_4}}{\partial \boldsymbol{f^U}} = \frac{\partial K(\bigstar^*,\blacklozenge^*)}{\partial \boldsymbol{f^U}} + \frac{\partial \mathcal{Q}(\bigstar^*,\boldsymbol{\gamma}, \boldsymbol{\mu})}{\partial \boldsymbol{f^U}}=0,\\
    \frac{\partial L_{\mathbb{P}_4}}{\partial \boldsymbol{f^E}} = \frac{\partial K(\bigstar^*,\blacklozenge^*)}{\partial \boldsymbol{f^E}} + \frac{\partial \mathcal{Q}(\bigstar^*,\boldsymbol{\gamma}, \boldsymbol{\mu})}{\partial \boldsymbol{f^E}}=0. \label{p4_L_K_Q_2}
\end{align}
\end{subequations}
\end{small}
Substituting (\ref{zhao_82_1})~--~(\ref{zhao_82_2}) into (\ref{p4_L_K_Q_1})~--~(\ref{p4_L_K_Q_2}), 
we can obtain:
\begin{small}
\begin{subequations}
\begin{align}
    \frac{\partial H(\bigstar^*)}{\partial \boldsymbol{\alpha}} + \frac{\partial \mathcal{Q}(\bigstar^*,\boldsymbol{\gamma}, \boldsymbol{\mu})}{\partial \boldsymbol{\alpha}}=0,\label{zhao_86a_1}\\ 
    \frac{\partial H(\bigstar^*)}{\partial \boldsymbol{p}} + \frac{\partial \mathcal{Q}(\bigstar^*,\boldsymbol{\gamma}, \boldsymbol{\mu})}{\partial \boldsymbol{p}}=0,\\
    \frac{\partial H(\bigstar^*)}{\partial \boldsymbol{b}} + \frac{\partial \mathcal{Q}(\bigstar^*,\boldsymbol{\gamma}, \boldsymbol{\mu})}{\partial \boldsymbol{b}}=0,\\
    \frac{\partial H(\bigstar^*)}{\partial \boldsymbol{f^U}} + \frac{\partial \mathcal{Q}(\bigstar^*,\boldsymbol{\gamma}, \boldsymbol{\mu})}{\partial \boldsymbol{f^U}}=0,\\
    \frac{\partial H(\bigstar^*)}{\partial \boldsymbol{f^E}} + \frac{\partial \mathcal{Q}(\bigstar^*,\boldsymbol{\gamma}, \boldsymbol{\mu})}{\partial \boldsymbol{f^E}}=0. \label{zhao_86a_2}
\end{align}
\end{subequations}
\end{small}
At the same time, with $\boldsymbol{\gamma}, \boldsymbol{\mu}$ denoting the Lagrange multipliers, the Lagrangian function of problem $\mathbb{P}_3$ can be given by:
\begin{align}
    L_{\mathbb{P}_3}(\bigstar,\blacklozenge,\boldsymbol{\gamma}, \boldsymbol{\mu}) = H(\bigstar)+\mathcal{Q}(\bigstar,\boldsymbol{\gamma}, \boldsymbol{\mu}).
\end{align}
Therefore, (\ref{zhao_86a_1})~--~(\ref{zhao_86a_2}) indicate that
\begin{align}
    \frac{\partial L_{\mathbb{P}_3}}{\partial \boldsymbol{\alpha}}=\frac{\partial L_{\mathbb{P}_3}}{\partial \boldsymbol{p}}=\frac{\partial L_{\mathbb{P}_3}}{\partial \boldsymbol{b}}=\frac{\partial L_{\mathbb{P}_3}}{\partial \boldsymbol{f^U}}=\frac{\partial L_{\mathbb{P}_3}}{\partial \boldsymbol{f^E}}=0.
\end{align}
Then, (\ref{p4_kkt_1})~--~(\ref{p4_kkt_4}) are equivalent to:
\\[-5pt]
\begin{numcases}{}
\text{Stationarity:}&  \nonumber\\
$$\frac{\partial L_{\mathbb{P}_3}}{\partial \boldsymbol{\alpha}}=\frac{\partial L_{\mathbb{P}_3}}{\partial \boldsymbol{p}}=\frac{\partial L_{\mathbb{P}_3}}{\partial \boldsymbol{b}}=\frac{\partial L_{\mathbb{P}_3}}{\partial \boldsymbol{f^U}}=\frac{\partial L_{\mathbb{P}_3}}{\partial \boldsymbol{f^E}}=0$$,  \label{p3_kkt_1} & \\
\text{Primal feasibility:}&  \nonumber \\
$$ \bigstar^*\text{~satisfy (\ref{P3_C_alpha}),~(\ref{P1_C_pn})~--~(\ref{P1_C_fnm_2})}$$,\label{p3_kkt_2}&\\
\text{Dual feasibility:}&  \nonumber \\
$$\gamma_{1,n}, \gamma_{2,n}, \gamma_{3,n}, \gamma_{4,n} \geq 0, \forall n \in \mathcal{N}$$,\label{p3_kkt_3} &\\
\text{Complementary slackness:} &  \nonumber\\
$$\gamma_{1,n}(1-\alpha_n^*)=0,  \gamma_{3,n} (p_n^*-p_n^{max}) =0$$,\nonumber & \\ 
$$\gamma_{2,n}(\alpha_n^*-\!\Upsilon) =0, \gamma_{4,n}(f_n^*-f_n^{max})=0, \forall n \in \mathcal{N}$$.&\label{p3_kkt_4} 
\end{numcases}
\\[+1pt]
The above (\ref{p3_kkt_1})~--~(\ref{p3_kkt_4}) indicate that $\bigstar^*$ is a stationary point for problem $\mathbb{P}_3$.
Therefore, the proof is concluded.
\end{proof}\vspace{-8pt}
It can be easily verified that $\mathbb{P}_4$ is a convex optimization problem and can be solved by utilizing convex optimization solvers such as CVX~\cite{grant2008cvx}.
According to Proposition \ref{proposition1}, we are able to obtain a stationary point for $\mathbb{P}_3$ after solving $\mathbb{P}_4$.

\vspace{-2pt}\subsection{Optimizing $\boldsymbol{\chi}$ given $\boldsymbol{\alpha,p,b,f^U,f^E}$}\vspace{-1pt}
Firstly, to reduce the computational complexity, we convert discrete variables into continuous ones.
Without loss of equivalence, constraint (\ref{P1_C_chi_1}) can be reformulated as:
\begin{numcases}{}
    $$\chi_{n,m} \in  [0,1],~\forall n \in \mathcal{N},~m \in \mathcal{M}$$, \label{P1_C_chi_1_1}\\
    $$\sum_{n \in \mathcal{N}} \sum_{m \in \mathcal{M}}~\chi_{n,m}\cdot(1-\chi_{n,m}) \leq 0. $$\label{P1_C_chi_1_2}
\end{numcases}
By replacing constraint (\ref{P1_C_chi_1}) with constraints (\ref{P1_C_chi_1_1}) and (\ref{P1_C_chi_1_2}), the discrete variables are transformed into continuous ones, thereby reducing the computation complexity of the problem. 

With fixed $\boldsymbol{\alpha,p,b,f^U,f^E}$, solving problem $\mathbb{P}_1$ is equivalent to solving the following problem:
\begin{subequations}\label{P5}
\begin{align}
    \text{Problem}~\mathbb{P}_5&(\boldsymbol{\alpha,p,b,f^U,f^E}):\nonumber\\&\min_{\boldsymbol{\chi}} \sum_{n \in \mathcal{N}} Cost_{n}^{u} + \sum_{m \in \mathcal{M}} Cost_{m}^{E}\tag{\ref{P5}},\nonumber\\
    &\text{~~~s.t.~(\ref{P1_C_chi_2}),~(\ref{P1_C_bn_2}),~(\ref{P1_C_fnm_2}),~(\ref{P1_C_chi_1_1}),~(\ref{P1_C_chi_1_2}).}\nonumber\\[-16pt]\nonumber
\end{align}
\end{subequations}
However, constraint (\ref{P1_C_chi_1_2}) remains a non-convex constraint.
Thus, further measures are required to efficiently tackle this challenge.
Next, we convert problem $\mathbb{P}_5$ into an equivalent problem that has linear constraints, which we then address using the CCCP method. 
To this end, we introduce the following lemma:
\begin{lem}
    Let $G(\chi_{n,m})=\sum_{n \in \mathcal{N}} Cost_{n}^{u} + \sum_{m \in \mathcal{M}} Cost_{m}^{E}$.  
    With any $\chi_{n,m}^{0}$ satisfying ~(\ref{P1_C_chi_2}),~(\ref{P1_C_bn_2}),~(\ref{P1_C_fnm_2}), and (\ref{P1_C_chi_1_1}), for all $\varrho > \varrho_0$ where
    \begin{align}
        \varrho_0 = \frac{G(\chi_{n,m}^0)-\min\{G(\chi_{n,m})\text{: (\ref{P1_C_chi_2}),~(\ref{P1_C_bn_2}),~(\ref{P1_C_fnm_2}),~(\ref{P1_C_chi_1_1})}\}}{\min\{ \sum_{n \in \mathcal{N}} \sum_{m \in \mathcal{M}} \chi_{n,m} (1\!-\!\chi_{n,m})\text{: (\ref{P1_C_chi_2}),~(\ref{P1_C_bn_2}),~(\ref{P1_C_fnm_2}),~(\ref{P1_C_chi_1_1})}\}},
    \end{align}
    problem $\mathbb{P}_5$ has the same optimal solution with problem $\mathbb{P}_6$, which is defined as follows:
    \begin{align}
    &\text{Problem}~\mathbb{P}_6(\boldsymbol{\alpha,p,b,f^U,f^E}):\nonumber\\&\min_{\boldsymbol{\chi}} \sum_{n \in \mathcal{N}} Cost_{n}^{u} + \!\!\sum_{m \in \mathcal{M}} Cost_{m}^{E}+  \varrho \!\sum_{n \in \mathcal{N}} \sum_{m \in \mathcal{M}} \!\chi_{n,m} (1-\chi_{n,m}),\\
    &\text{~~~s.t.~(\ref{P1_C_chi_2}),~(\ref{P1_C_bn_2}),~(\ref{P1_C_fnm_2}),~(\ref{P1_C_chi_1_1}).}\nonumber
\end{align}
It is worth noting that problem $\mathbb{P}_6$ is derived from problem $\mathbb{P}_5$ by integrating the concave constraint (\ref{P1_C_chi_1_2}) into the objective function as a penalization term.
\end{lem}
\vspace{-3pt}\begin{proof}
    The proof can be directly derived from Theorem 1 in~\cite{an2012exact}.
\end{proof}\vspace{-10pt}

Problem $\mathbb{P}_6$ involves subtracting a quadratic convex function from a linear function, while its constraints are linear in nature.
According to~\cite{thi1997solving}, problem $\mathbb{P}_6$ falls under the category of indefinite quadratic problem, which is a subset of the broader class of problems known as the difference of convex problems.
With the objective function of problem $\mathbb{P}_6$ being differentiable, we can effectively address problem $\mathbb{P}_6$ using the CCCP method, which involves employing a first-order Taylor series approximation~\cite{hanif2015minorization} to refine $\sum_{n \in \mathcal{N}} \sum_{m \in \mathcal{M}}\chi_{n,m} (\chi_{n,m}-1)$.
Specifically, it updates the expression to:
\begin{align}
    \sum_{n \in \mathcal{N}} \sum_{m \in \mathcal{M}} \!\!\chi_{n,m}^{(i)} (\chi_{n,m}^{(i)}\!-\!1)+\!\!\sum_{n \in \mathcal{N}} \sum_{m \in \mathcal{M}}\!\! (2\chi_{n,m}^{(i)}\!-\!1)(\chi_{n,m}\!-\!\chi_{n,m}^{(i)}),
\end{align}
where $\chi_{n,m}^{(i)}$ indicates the value of $\chi_{n,m}$ at the $i$-th iteration.
After that, the CCCP method systematically approaches resolution by iteratively engaging in a sequence of linear problems. 
The CCCP method not only simplifies complex issues by breaking them down into more manageable linear tasks but also ensures a structured progression towards finding an optimal solution through successive approximations.
However, directly solving problem $\mathbb{P}_6$ to reach a feasible solution for problem $\mathbb{P}_5$ might not always be viable. 
To navigate this challenge, our approach involves generating several local optimum solutions for problem $\mathbb{P}_6$. 
This is achieved by applying the CCCP algorithm multiple times, initiating each iteration from a different feasible starting point specific to problem $\mathbb{P}_6$. 
The optimal solution is then determined by selecting the one that presents the smallest average value among these.

\section{Simulations}\label{sec-Simulation}

In this section, we present the performance of the proposed approach through simulations.
The simulated MEC network has 50 mobile users and 10 edge servers by default.
Assume the users and edge servers collaboratively train  Meta's open-source large language model Meta-AI (LLaMA-7B) which consists of 32 transformer layers~\cite{touvron2023llama2}.
The path loss is modeled as $128.1 + 37.6\log(\texttt{distance})$ and Gaussian noise power is $\sigma^2=-134\text{dBm}$.
The maximum transmission power $p_n^{max}$ for the users is set in the range of 1 to 2 W.
The maximum GPU frequency $f_n^{max}$ for users and $f_m^{max}$ for edge servers are chosen from [0.5,1] and [1,3] respectively.
The total bandwidth $b_m^{max}$ for each edge server is 20 MHz.

For LLM training, the batch size $B$ is set to 512 and the dimensionality of the hidden states $h$ is set to 1024.
The lengths of input tokens for each user are randomly generated from 512 to 1024.
We assume the mobile users are equipped with mobile devices with GPU such as Apple A15, whose GPU has 4 to 6 cores.
Thus, the number of cores of the GPU at the user side $C_n^U$ is chosen between 4 to 6.
The number of FLOPs per cycle per core $D_n^U$ is all set to 1.
The edge servers are presumed to be equipped with advanced GPUs such as NVIDIA Tesla T4 and NVIDIA Tesla V100, therefore the number of cores of the GPU $C_m^E$ is randomly assigned values from the interval [2560, 5120].
The number of FLOPs per cycle per core $D_m^E$ is chosen between 1 and 2.

\subsection{The performance of the proposed collaborative training method}
\begin{figure}[t]
\captionsetup[subfigure]{justification=centering}
  \centering
  \captionsetup{justification=centering}
    \subfigure[The system performance on energy consumption.]{              
        \includegraphics[width=1.5in]{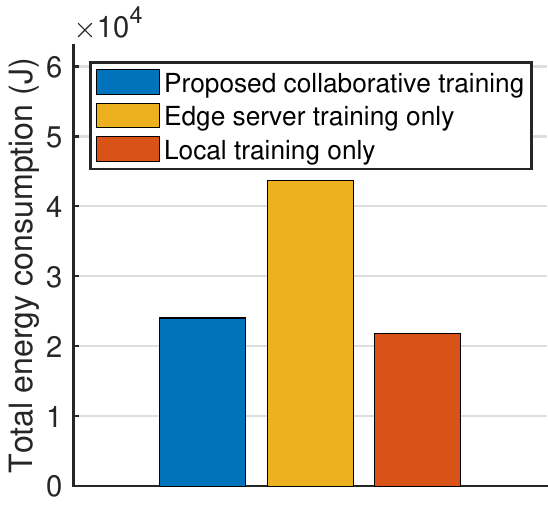}}
    \hfill
    \subfigure[The system performance on average delay.]{
        \includegraphics[width=1.5in]{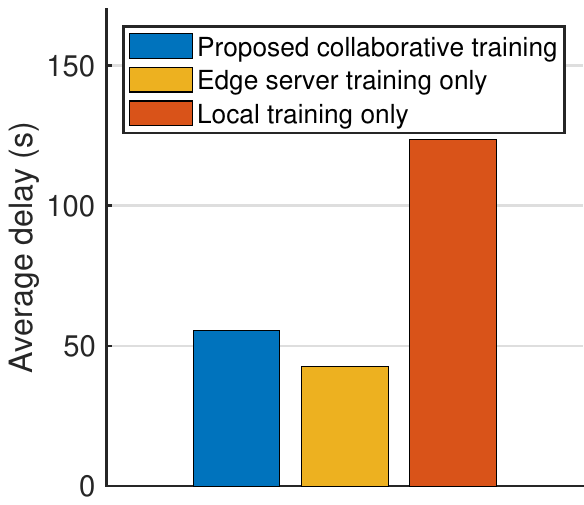}}
    \vspace{-14pt}\caption{Comparison of system performance with and without the proposed collaborative training approach.}\label{fig_before_after_collaborative_training}\vspace{-18pt}
\end{figure}

In Figure \ref{fig_before_after_collaborative_training}, we present an analysis of system performance across three computing approaches: the proposed collaborative training method, edge server training method and local training method. 
Figure \ref{fig_before_after_collaborative_training} (a) illustrates the energy consumption associated with each approach, where the proposed collaborative method demonstrates a balanced reduction in energy usage compared to the edge server training and local training methods. 
Figure \ref{fig_before_after_collaborative_training} (b) depicts the delay experienced under each approach, showing that the proposed method effectively minimizes delay, achieving a performance closer to the edge server training approach while significantly outperforming the local training method. 
These results demonstrate the efficiency and effectiveness of the proposed collaborative training method in optimizing both energy consumption and system delay.

\subsection{The performance of proposed algorithms under weighting factors}

\begin{figure*}
\vspace{-5pt}
\begin{minipage}{0.73\linewidth}
\captionsetup[subfigure]{justification=centering}
  \centering
  \captionsetup{justification=centering}
    \subfigure[The total energy consumption under different weighting factors for energy.]{\includegraphics[width=0.32\textwidth]{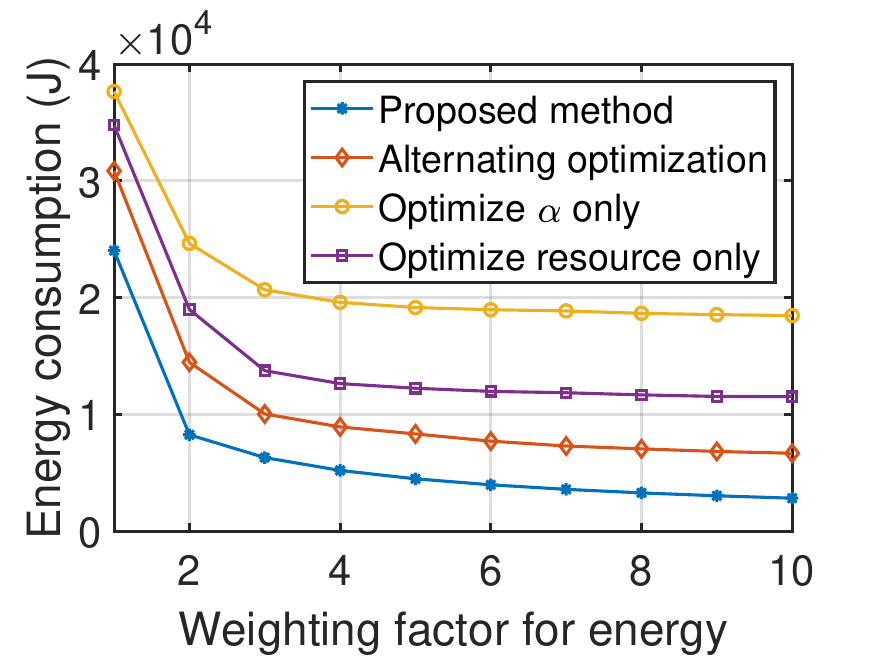}}
    \hfill
    \subfigure[The average delay under different weighting factors for energy.]{\includegraphics[width=0.32\textwidth]{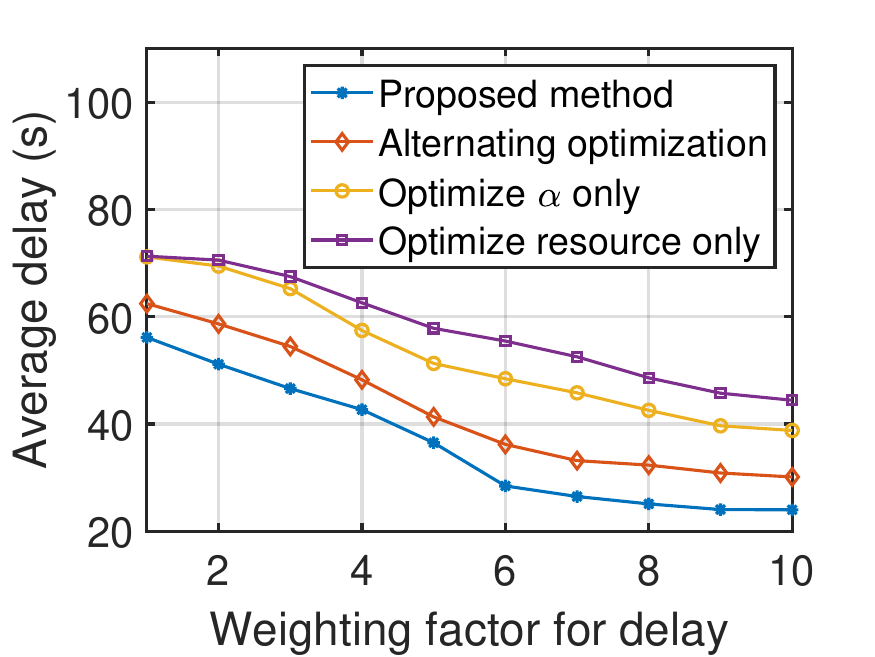}}
     \hfill
    \subfigure[The average model stability under different weighting factors for model stability.]{\includegraphics[width=0.32\textwidth]{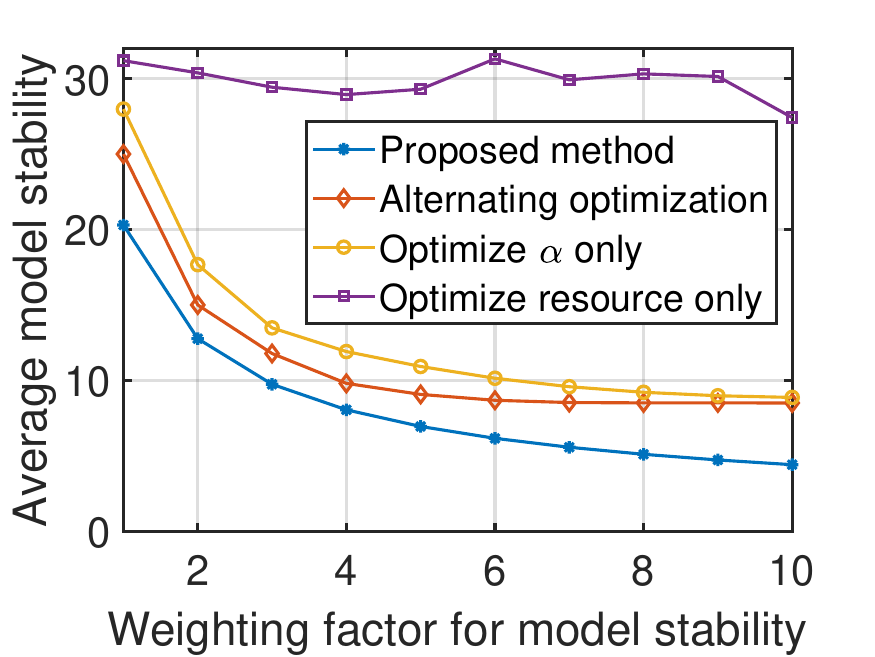}}
    \vspace{-15pt}\caption{The performance of the proposed method under different weighting factors.}\label{fig_weighting_factor}
\end{minipage}
\begin{minipage}{0.25\linewidth}
\centering
\vspace{+5pt}\includegraphics[width=0.95\textwidth]{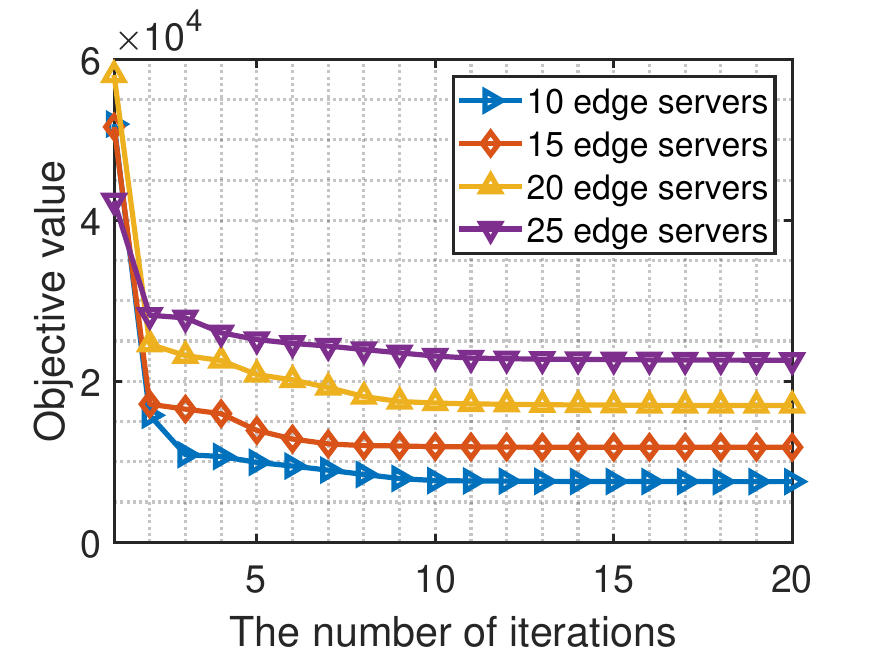} 
\vspace{-7pt}\caption{The convergence performance with different numbers of edge servers.}
\label{fig_convergence}
\end{minipage}\vspace{-8pt}
\end{figure*}
Next, we compare the performance of the proposed method when the weighting factors for energy, delay and model stability vary.
The default weighting factors after normalization are all set to 1 for energy, delay and model stability.
A larger weighting factor denotes enhanced prioritization of system attributes such as energy efficiency, latency, or model stability.
The three additional methodologies employed for comparative analysis with our proposed method are listed as follows:
\begin{itemize}[leftmargin=*]
  \item \textbf{Alternating optimization:} This method is the most commonly employed strategy in the related literature as discussed in Section~\ref{sec-Related-Work}. 
  It systematically alternates between optimizing offloading decisions and the allocation of computational or communication resources.
  \item \textbf{Optimize $\alpha$ only:} This approach solely focuses on optimizing the offloading decision $\alpha$, while implementing a random strategy for resource allocation.
  \item \textbf{Optimize resource only:} This method concentrates exclusively on the optimization of resource allocation, while employing a random approach to the offloading decision $\alpha$.
\end{itemize}
In Figure~\ref{fig_weighting_factor}, we adjust the weighting factors for energy, delay and model stability from 1 to 10, respectively.
For each setting where one attribute's weighting factor varied from 1 to 10, the weighting factors for the other two attributes are held constant at 1.
In Figure~\ref{fig_weighting_factor} (a), the proposed methodology consistently attains the lowest energy consumption among the methods evaluated.
The alternating optimization approach secures the second-best performance. Conversely, the method that solely optimizes $\alpha$ exhibits the poorest results. 
This suboptimal performance can be attributed to the fact that the $\alpha$-only optimization method neglects resource allocation considerations, which are crucial in minimizing energy consumption. 
Furthermore, as the weighting factor for energy is incrementally increased, the reductions in optimal energy consumption diminish progressively, eventually converging.
Figure~\ref{fig_weighting_factor} (b) depicts the average delay experienced under various weighting factors for delay.
Consistently, the proposed approach yields the minimal delay, surpassing the performance of the alternating optimization method.
Notably, the strategy focusing solely on optimizing $\alpha$ demonstrates superior results compared to that which exclusively optimizes resource allocation.
This advantage is attributed to the enhanced computational capabilities of the edge servers, which significantly reduce computational delays.
Figure~\ref{fig_weighting_factor} (c) illustrates the mean stability of the model across a range of weighting factors for model stability. 
The method we proposed consistently attains the highest level of model stability.
The alternating optimization approach outperforms the strategy that solely optimizes $\alpha$ a little, although both methods converge to nearly identical point in the long term. 
Conversely, the technique that focuses exclusively on optimizing resource allocation exhibits the poorest performance, primarily due to the arbitrary selection of the offloading decision $\alpha$, which significantly impacts model stability.

\subsection{The impact of the number of users and the number of edge servers}
\begin{figure}
\captionsetup[subfigure]{justification=centering}
  \centering
  \captionsetup{justification=centering}
    \subfigure[The energy consumption under different numbers of users.]{              
        \includegraphics[width=1.6in]{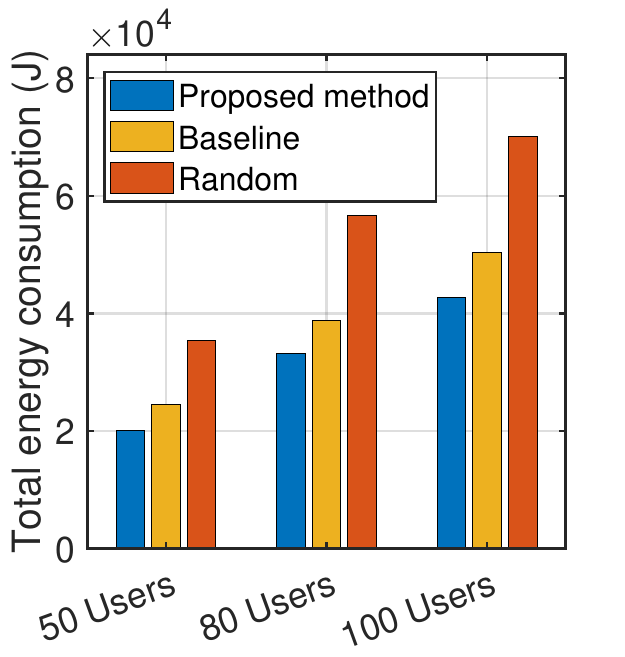}}
    \hfill
    \subfigure[The average delay under different numbers of users.]{
        \includegraphics[width=1.6in]{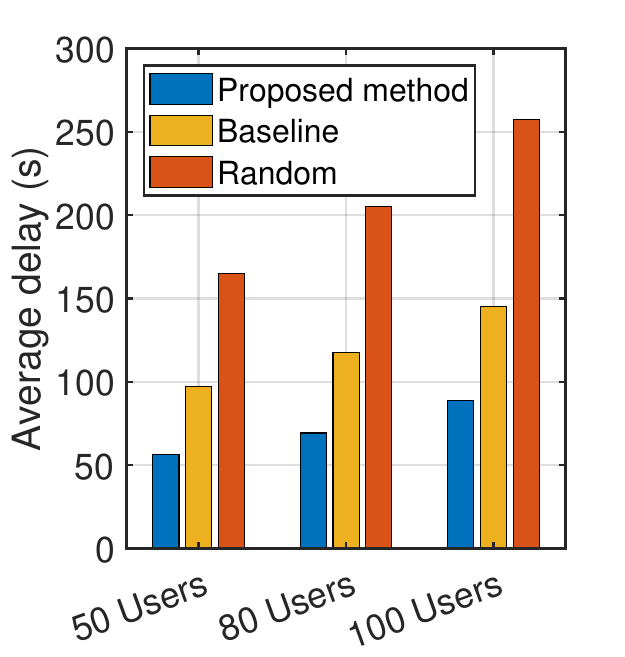}}
    \vspace{-11pt}\caption{The performance of the algorithm under different numbers of mobile users.}\label{fig_number_of_uses}\vspace{-15pt}
\end{figure}

In this part, we assess the influence of both the user population and the number of edge servers on the effectiveness of the proposed approach to addressing the user-to-edge association challenge. 
We employ a comparative methodology as outlined below:
\begin{itemize}[leftmargin=*]
  \item \textbf{Baseline:} The baseline approach we choose is a greedy-based strategy. Under this strategy, each user opts for the edge server offering the highest available transmission rate, subject to bandwidth limitations.
  \item \textbf{Random:} The random user-to-edge server association method distributes users among edge servers in a stochastic manner, also adhering to bandwidth constraints.
\end{itemize}
In Figure~\ref{fig_number_of_uses} (a), we present the total energy consumption when there are different numbers of users.
It can be observed that the proposed method always outperforms the two alternative strategies.
The baseline strategy selects the edge server with the highest available transmission rate for each user; however, this approach may inadvertently overload servers that possess lower computational efficiency, thereby causing a marginal increase in energy consumption relative to our method. 
The random strategy invariably results in the highest energy expenditure.
Subsequently, we analyze the average delay contingent on varying user quantities, as depicted in Figure~\ref{fig_number_of_uses} (b).
It is evident that the proposed methodology consistently surpasses the two alternative strategies.
Specifically, the random strategy yields the least favorable outcomes due to its arbitrary selection of edge servers for user allocation.
While the baseline strategy may attain minimal communication delays, it tends to allocate an excessive number of users to a single edge server, thereby exacerbating the computational delays.

In Figure~\ref{fig_convergence}, the convergence performance of the algorithm is analyzed for the user-to-edge server association problem across varying quantities of edge servers. 
For each scenario considered, the user count remains constant at 100. The analysis reveals that the algorithm attains a stationary point as evidenced by the stabilization of the objective value. 
It is worth noticing that configurations with a smaller number of edge servers exhibit faster convergence rates. 
This enhanced speed of convergence can be attributed to the diminished complexity of the optimization challenge: a reduced number of servers correlates with fewer constraints and a lower number of parameters requiring adjustment throughout the optimization procedure. 
In all tested configurations, the algorithm consistently achieved convergence within 13 iterations, thereby demonstrating its robust capability to efficiently resolve the user-to-edge server association problem.


\vspace{-6pt}\section{Conclusion}\vspace{-3pt}\label{sec-Conclusion}
In this study, we present a collaborative LLM training framework that merges the efforts of mobile users and edge servers. 
Here, mobile users are tasked with training the preliminary layers of the LLM, while the computationally intensive later layers are managed by the edge servers. 
We develop a multi-objective optimization strategy aimed at reducing overall energy usage and latency experienced by users, while also improving the stability of LLMs. 
Through analytical analysis, we establish an upper bound for the average-replace-one stability. 
The proposed algorithm leverages fractional programming and the CCCP method to derive solutions. 
Simulation results indicate that our approach effectively reduces energy usage and delay, and enhances the stability of LLMs in the mobile edge computing environments.

\vspace{-6pt}\section*{Acknowledgement}\vspace{-3pt}

This research is supported by the National Research Foundation, Singapore and Infocomm Media Development Authority under its Trust Tech Funding Initiative, Singapore MOE AcRF Tier 1 RT5/23, Tier 1 RG90/22, and NTU-WASP Joint Project. Any opinions, findings and conclusions or recommendations expressed in this material are those of the author(s) and do not reflect the views of National Research Foundation, Singapore and Infocomm Media Development Authority.



\begin{appendix}
\setcounter{equation}{0}
\renewcommand\theequation{A.\arabic{equation}}
\section{Proof of Theorem \ref{theorem_AS}}\label{Appendix.A}
Denote $\textbf{w}$ as the parameters of a language model, and $\xi=\dim(\textbf{w})$ as the number of the parameters. 
When the users fine-tune a pre-trained model, denoting $M^{\xi \times \xi}$ as a mask matrix which is a diagonal matrix with $M_{ii}=\{0,1\}$ and $\mathcal{L}_\mathcal{S}(\textbf{w})=\sum_{i=1}^k \ell(\textbf{w},z_i)$ for a training dataset $\mathcal{S}$, the training process of fine-tuning the pre-trained model is to solve the following problem:
\begin{subequations}\label{FuZihao_(2)}
\vspace{-3pt}\begin{align}
    &\min_{\textbf{w}}~ \mathcal{L}_\mathcal{S}(\textbf{w}),\tag{\ref{FuZihao_(2)}}\\[-3pt]
    \text{s.t.~}\Vert(I-&M)(\textbf{w}-\textbf{w}^0)\Vert^2=0,\\[-17pt]\nonumber
\end{align}
\end{subequations}
where $\textbf{w}^0$ are the parameters of the pre-trained model.
According to the Lagrangian duality, problem (\ref{FuZihao_(2)}) is equivalent to:
\begin{small}
\vspace{-2pt}\begin{align}\label{FuZihao_(3)}
     \min_\textbf{w}~\max_\lambda~ \mathcal{L}_\mathcal{S}(\textbf{w}) + \lambda\Vert(I-M)(\textbf{w}-\textbf{w}^0)\Vert^2,\\[-16pt]\nonumber
\end{align} 
\end{small}
where $\lambda \geq 0$ is the Lagrangian multiplier.
Since
\begin{small}
\vspace{-3pt}\begin{align}
\min_\textbf{w}~\max_\lambda~\mathcal{L}_\mathcal{S}(\textbf{w}) + \lambda\Vert(I-M)(\textbf{w}-\textbf{w}^0)\Vert^2 \nonumber \geq \\[-3pt] \min_\textbf{w}~ \mathcal{L}_\mathcal{S}(\textbf{w}) + \Vert(I-M)(\textbf{w}-\textbf{w}^0)\Vert^2,\label{11}\\[-16pt]\nonumber
\end{align} 
\end{small}
we then focus on optimizing the lower bound of problem (\ref{FuZihao_(2)}), which is given by:
\begin{small}
\vspace{-2pt}\begin{align}\label{FuZihao_(4)}
    \min_\textbf{w}~ \mathcal{L}_\mathcal{S}^\prime(\textbf{w}) = \mathcal{L}_\mathcal{S}(\textbf{w}) + \Vert(I-M)(\textbf{w}-\textbf{w}^0)\Vert^2.\\[-16pt]\nonumber
\end{align} 
\end{small}
It indicates that minimizing initial loss function $\mathcal{L}_\mathcal{S}(\textbf{w})$, augmented by the regularization term $\Vert(I-M)(\textbf{w}-\textbf{w}^0)\Vert^2$ provides a lower bound on the optimal value of problem (\ref{FuZihao_(2)}).

By taking the expectation with respect to $M$, we can get:
\begin{small}
\vspace{-4pt}\begin{align}
\mathbb{E}_M\Big(\mathcal{L}_\mathcal{S}^\prime(\textbf{w})\Big) &=  \mathcal{L}_\mathcal{S}(\textbf{w}) + \mathbb{E}\Vert(I-M)(\textbf{w}-\textbf{w}^0)\Vert^2\nonumber\\[-4pt]
    &=\mathcal{L}_\mathcal{S}(\textbf{w}) +  \Vert(\textbf{w}-\textbf{w}^0)\Vert^2\mathbb{E}\Big(\sum_{i=1}^\xi(1-M_{ii})^2\Big)\nonumber\\[-2pt]
    &=\mathcal{L}_\mathcal{S}(\textbf{w})+\Vert(\textbf{w}-\textbf{w}^0)\Vert^2(1-\alpha),
\end{align}
\end{small}
where the validity of the last equality is attributed to the fact that the fraction of parameters subjected to fine-tuning is $\alpha$.
Therefore, $\mathcal{A}(\mathcal{S})$ can be given by:
\vspace{-4pt}\begin{align}
    \mathcal{A}(\mathcal{S})=\argmin_{\textbf{w}}~\mathcal{L}_\mathcal{S}(\textbf{w})+(1-\alpha)\Vert\textbf{w}-\textbf{w}^0\Vert^2.\\[-17pt]\nonumber
\end{align}
Next, we denote $f_\mathcal{S}(\textbf{w})=\mathcal{L}_\mathcal{S}(\textbf{w})+(1-\alpha)\Vert\textbf{w}-\textbf{w}^0\Vert^2$.
Subsequently, $\forall \textbf{u},\textbf{v}, \forall i\in\{1,\ldots,k\}$, we can get:
\begin{small}
\vspace{-4pt}\begin{align}\label{13.8}
    &f_\mathcal{S}(\textbf{u})-f_\mathcal{S}(\textbf{v})\nonumber\\[-4pt] &=\mathcal{L}_\mathcal{S}(\textbf{u})\!+\!(1\!-\!\alpha)\Vert\textbf{u}\!-\!\textbf{u}^0\Vert^2 - \Big( \mathcal{L}_\mathcal{S}(\textbf{v})\!+\!(1\!-\!\alpha)\Vert\textbf{v}\!-\!\textbf{v}^0\Vert^2\Big) \nonumber \\[-2pt]
    &=\mathcal{L}_{\mathcal{S}^{i}}(\textbf{u})\!+\!(1\!-\!\alpha)\Vert\textbf{u}\!-\!\textbf{u}^0\Vert^2 - \Big( \mathcal{L}_{\mathcal{S}^{i}}(\textbf{v})\!+\!(1\!-\!\alpha)\Vert\textbf{v}\!-\!\textbf{v}^0\Vert^2\Big) \nonumber \\[-2pt]
    &~~~~~~~~~~~~~~~~~~~~-\frac{\ell(\textbf{u},z_i^\prime)\!-\!\ell(\textbf{v},z_i^\prime)}{k}\!+\!\frac{\ell(\textbf{u},z_i)\!-\!\ell(\textbf{v},z_i)}{k}\nonumber\\[-2pt]
    &=f_{\mathcal{S}^{i}}(\textbf{u})\!-\!f_{\mathcal{S}^{i}}(\textbf{v})\!-\!\frac{\ell(\textbf{u},z_i^\prime)\!-\!\ell(\textbf{v},z_i^\prime)}{k}\!+\!\frac{\ell(\textbf{u},z_i)\!-\!\ell(\textbf{v},z_i)}{k}.\\[-17pt]\nonumber
\end{align}
\end{small}
Let's set $\textbf{u}=\mathcal{A}(\mathcal{S}^{i})$ and $\textbf{v}=\mathcal{A}(\mathcal{S})$ in (\ref{13.8}).
Using the fact that $f_{\mathcal{S}^{i}}(\textbf{u})\leq f_{\mathcal{S}^{i}}(\textbf{v})$ because $\mathcal{A}(\mathcal{S}^{i})$ is the minimizer of $f_{\mathcal{S}^{i}}(\textbf{w})$,
we can get:
\begin{small}
\begin{align}\label{13.9}
    &f_\mathcal{S}(\mathcal{A}(\mathcal{S}^{i}))-f_\mathcal{S}(\mathcal{A}(\mathcal{S})) \leq\nonumber\\& \frac{\ell(\mathcal{A}(\mathcal{S}^{i}),z_i)-\ell(\mathcal{A}(\mathcal{S}),z_i)}{k}-\frac{\ell(\mathcal{A}(\mathcal{S}^{i}),z_i^\prime)-\ell(\mathcal{A}(\mathcal{S}),z_i^\prime)}{k}.\\[-15pt]\nonumber
\end{align}
\end{small}
Since the loss function is strong convex, $f_\mathcal{S}$ is $2(1-\alpha)$-strong convex, which means $\forall \textbf{u},\textbf{v}$,
\begin{small}
\vspace{-2pt}\begin{align}\label{strongly_convex}
    f_\mathcal{S}(\textbf{u}) \geq f_\mathcal{S}(\textbf{v})+\nabla f_\mathcal{S}(\textbf{v})^\intercal(\textbf{u}-\textbf{v})+(1-\alpha) \Vert\textbf{u}-\textbf{v}\Vert^2.\\[-17pt]\nonumber
\end{align}
\end{small}
Let $\textbf{u}=\mathcal{A}(\mathcal{S}^{i})$ and $\textbf{v}=\mathcal{A}(\mathcal{S})$, then $\nabla f_{\mathcal{S}}(\mathcal{A}(\mathcal{S}))=0$ since $\mathcal{A}(\mathcal{S})$ is the minimizer of $f_{\mathcal{S}}(\textbf{w})$.
Therefore, (\ref{strongly_convex}) becomes:
\begin{small}
\vspace{-3pt}\begin{align}\label{13.7}
f_\mathcal{S}(\mathcal{A}(\mathcal{S}^{i})) \geq f_\mathcal{S}(\mathcal{A}(\mathcal{S}))+(1-\alpha) \Vert\mathcal{A}(\mathcal{S}^{i})-\mathcal{A}(\mathcal{S})\Vert^2,\\[-15pt]\nonumber
\end{align}
\end{small}
which can be rearranged as follows:
\begin{small}
\vspace{-3pt}\begin{align}\label{13.7-2}
f_\mathcal{S}(\mathcal{A}(\mathcal{S}^{i}))-f_\mathcal{S}(\mathcal{A}(\mathcal{S})) \geq (1-\alpha) \Vert\mathcal{A}(\mathcal{S}^{i})-\mathcal{A}(\mathcal{S})\Vert^2.\\[-15pt]\nonumber
\end{align}
\end{small}
Combing (\ref{13.9}) and (\ref{13.7-2}), it yields:
\begin{small}
\vspace{-3pt}\begin{align}\label{13.10}
    &(1-\alpha) \Vert\mathcal{A}(\mathcal{S}^{i})-\mathcal{A}(\mathcal{S})\Vert^2 \leq \nonumber\\
    &\frac{\ell(\mathcal{A}(\mathcal{S}^{i}),z_i)-\ell(\mathcal{A}(\mathcal{S}),z_i)}{k}-\frac{\ell(\mathcal{A}(\mathcal{S}^{i}),z_i^\prime)-\ell(\mathcal{A}(\mathcal{S}),z_i^\prime)}{k}.\\[-15pt]\nonumber
\end{align}
\end{small}
Since we assume $\ell(\cdot,z_i)$ is $L$-Lipschitz, which means:
\vspace{-3pt}\begin{small}
\begin{align}\label{13.11_1}
    \ell(\mathcal{A}(\mathcal{S}^{i}), z_i) - \ell(\mathcal{A}(\mathcal{S}), z_i)\leq L\Vert\mathcal{A}(\mathcal{S}^{i})- \mathcal{A}(\mathcal{S})\Vert,\\[-17pt]\nonumber
\end{align} 
\end{small}
and
\vspace{-3pt}\begin{small}
\begin{align}\label{13.11_2}
    \ell(\mathcal{A}(\mathcal{S}), z_i^\prime) - \ell(\mathcal{A}(\mathcal{S}^{i}), z_i^\prime)\leq L\Vert\mathcal{A}(\mathcal{S}^{i})- \mathcal{A}(\mathcal{S})\Vert.\\[-15pt]\nonumber
\end{align} 
\end{small}
Substituting (\ref{13.11_1}) and (\ref{13.11_2}) into (\ref{13.10}) yields the following result:
\vspace{-9pt}\begin{small}
\begin{align}
    (1-\alpha) \Vert\mathcal{A}(\mathcal{S}^{i})-\mathcal{A}(\mathcal{S})\Vert^2 \leq \frac{2L\Vert\mathcal{A}(\mathcal{S}^{i})- \mathcal{A}(\mathcal{S})\Vert}{k},\\[-17pt]\nonumber
\end{align} 
\end{small}
which indicates:
\vspace{-3pt}\begin{small}
\begin{align}\label{eq1}
    \Vert\mathcal{A}(\mathcal{S}^{i})-\mathcal{A}(\mathcal{S})\Vert \leq \frac{2L}{(1-\alpha) k}.\\[-17pt]\nonumber
\end{align} 
\end{small}
Inserting (\ref{eq1}) into (\ref{13.11_1}) leads us to determine the following:
\vspace{-3pt}\begin{small}
\begin{align}
    \ell(\mathcal{A}(\mathcal{S}^{i}), z_i) - \ell(\mathcal{A}(\mathcal{S}), z_i)\leq \frac{2L^2}{(1-\alpha) k}.\\[-17pt]\nonumber
\end{align} 
\end{small}
Given that this is true for any $\mathcal{S}$ and $z_i$, we can finally deduce that $\forall i \in \{1, \ldots, k\}$:
\vspace{-5pt}\begin{small}
\begin{align}\label{corollary_13.6}
    \mathbb{E}_{\mathcal{S}} \!\left[ \left| \ell(\mathcal{A}(\mathcal{S}), z_i) - \ell(\mathcal{A}(\mathcal{S}^{i}), z_i) \right| \right] \leq \frac{2L^2}{(1-\alpha) k}.\\[-17pt]\nonumber
\end{align} 
\end{small}
\end{appendix}

\vspace{-15pt}\bibliographystyle{ACM-Reference-Format}
\vspace{-6pt}\bibliography{related}

\end{document}